%% file: masterFile.tex
\documentclass[prl,twocolumn,floatfix,amsfonts]{revtex4}
\usepackage{graphicx,graphics,color,epsfig}
\usepackage{bm}
\usepackage{amssymb,amsfonts,amsmath}
\usepackage{epstopdf}

\def\Re{\text{Re}}
\def\Im{\text{Im}}

\begin{document}

\include{FKL}

\setcounter{figure}{0}
\setcounter{equation}{0}
\include{SI}

\end{document}

%% file: FKL.tex



\title{Metallic ferromagnetism in the Kondo lattice}

\author{Seiji J.~Yamamoto}
\affiliation{National High Magnetic Field Laboratory  and Florida State University, Tallahassee, FL 32310, USA}
\author{Qimiao Si}
\affiliation{Department of Physics and Astronomy, Rice University, Houston, Texas 77005, USA}

\begin{abstract}
Metallic magnetism is both ancient and modern,
occurring in such familiar settings as the lodestone in compass needles
and the hard drive in computers.
Surprisingly, a rigorous theoretical basis for metallic ferromagnetism
is still largely missing.
The Stoner approach perturbatively treats Coulomb
interactions when the latter need to be large,
while the Nagaoka approach incorporates thermodynamically 
negligible holes into a half-filled band.
Here, we show that the ferromagnetic order of the Kondo lattice
is amenable to an asymptotically exact analysis
over a range of interaction parameters.
In this ferromagnetic phase, the conduction electrons and local moments
are strongly coupled but the Fermi surface 
does not enclose the latter ({\it i.e.}, it is ``small'').
Moreover, non-Fermi liquid behavior appears over a 
range of frequencies and temperatures.
Our results provide the basis to understand some long-standing puzzles
in the ferromagnetic heavy fermion metals,
and raise the prospect for a new class of 
ferromagnetic quantum phase transitions.
\end{abstract}

\maketitle

A contemporary theme in quantum condensed matter physics 
concerns competing ground states and the accompanying novel
excitations~\cite{Coleman05}.
With a plethora of different phases, 
magnetic 
heavy fermion materials
should reign supreme as the prototype for competing
order.
So far, most of the theoretical scrutiny
has focused on antiferromagnetic
heavy fermions~\cite{Gegenwart08, HvL07}.
Nonetheless, the list of heavy fermion metals which are known 
to exhibit ferromagnetic order continues to grow. 
An early example subjected to extensive studies 
is CeRu$_{\rm 2}$Ge$_{\rm 2}$
(ref.~\cite{Sullow99} and references therein).
Other ferromagnetic heavy fermion metals include 
CePt~\cite{Larrea05},  CeSi$_{\rm x}$~\cite{Drotziger06}, 
CeAgSb$_{\rm 2}$~\cite{Sidorov03},
and URu$_{\rm 2-x}$Re$_{\rm x}$Si$_{\rm 2}$ 
at 
${\rm x} > 0.15$~\cite{Bauer05,Butch09}.
More recently discovered materials include CeRuPO~\cite{Krellner07}
and UIr$_{\rm 2}$Zn$_{\rm 20}$~\cite{Bauer06}.
Finally,
systems 
such as UGe$_{\rm 2}$~\cite{Saxena00} and URhGe~\cite{Levy07}
are particularly interesting
because they
exhibit a superconducting dome as 
their metallic ferromagnetism is tuned toward
its border.
Some fascinating and general questions
have emerged
~\cite{King91, Yamagami94, Ikezawa97},
yet they have hardly been addressed theoretically.
One central issue concerns the nature of the Fermi surface:
Is
it 
``large,'' encompassing both the local moments 
and conduction electrons as in paramagnetic heavy fermion
metals~\cite{Hewson97, Oshikawa00},
or is it ``small,''
incorporating only conduction electrons?
Measurements of 
the de Haas-van Alphen (dHvA) effect
have
suggested that the Fermi surface is 
small in 
CeRu$_{\rm 2}$Ge$_{\rm 2}$~\cite{King91, Yamagami94, Ikezawa97},
and have provided evidence for 
Fermi surface reconstruction as a function of pressure
in UGe$_{\rm 2}$~\cite{Settai02,Huxley}.
At the same time, 
it is traditional to consider the heavy fermion ferromagnets
as having a large Fermi surface when their relationship with
unconventional superconductivity is
discussed~\cite{Saxena00, Levy07,Schofield03};
an alternative form of the Fermi surface in the ordered state
could give rise to a new type of superconductivity
near its phase boundary.
All these point to the importance of theoretically understanding 
the ferromagnetic phases of heavy fermion metals,
and this will be the focus of 
the present work.

We 
consider 
the Kondo lattice model
in which 
a periodic array
of local moments interact
with each other
and with a conduction-electron band.
Kondo lattice systems are normally studied in the 
paramagnetic state,
where Kondo screening 
leads to heavy quasiparticles in the single-electron excitation
spectrum~\cite{Hewson97}.
The Stoner~\cite{Stoner38} mean field treatment of these heavy quasiparticles 
may then
lead to an itinerant ferromagnet~\cite{Perkins07}.
With the general limitations of the
Stoner approach in mind, here we carry out an asymptotically
exact analysis
of the ferromagnetic state.
We are able to do so by using a reference point that 
differs from either the Stoner or Nagaoka approach~\cite{Nagaoka66}, 
and accessing a ferromagnetic phase whose excitations 
are of considerable interest in the context of 
heavy fermion ferromagnets. We should stress that 
a ferromagnetic order may also arise in different regimes
of related models, such as in one dimension~\cite{Sigrist92} 
or in the presence of mixed-valency~\cite{Batista03}.

Our 
model contains a lattice of spin-$\frac{1}{2}$
local moments (${\bf S}_i$ for each site $i$) with a ferromagnetic
exchange interaction ($I<0$), a band of conduction electrons
($c_{{\vec{K}}\sigma}$, where ${\vec{K}}$ is the wavevector and $\sigma$
the spin index) with a dispersion $\epsilon_{\vec{K}}$ 
and a characteristic bandwidth $W$,
and an on-site antiferromagnetic Kondo exchange interaction ($J_K>0$)
between the local moments and the spin of the conduction electrons.
The corresponding Hamiltonian is
\begin{eqnarray}\label{eq:KLHamiltonian}
  H=
	\sum_{{\vec{K}}} \epsilon^{\phantom\dagger}_{\vec{K}}
c^{\dagger}_{{\vec{K}}\sigma} c^{\phantom\dagger}_{{\vec{K}}\sigma} + 
I \sum_{\langle ij \rangle} S_i^a S_j^a  
+  \sum_{i}
J_K^a S_i^a 
c^{\dagger}_{i\sigma} 
\frac{ \tau^a_{\sigma\sigma^{\prime}} }{2}
c^{\phantom\dagger}_{i\sigma^{\prime} } .
\end{eqnarray}
The symbol $\tau$ represents the Pauli matrices, 
with indices $a \in \{ x,y,z \}$ and $\sigma \in \{ \uparrow, \downarrow \}$.
Here $I$ represents the sum of direct exchange interaction between the
local moments and the effective exchange interaction generated by the 
conduction electron states that are not included in Eq. (1). Incorporating
this explicit exchange interaction term allows the study of the global
phase diagram of the Kondo lattice systems, and tuning a control
parameter in any specific heavy fermion material represents taking
a cut within this phase diagram. The Hamiltonian above is to be contrasted 
with models for double-exchange ferromagnets in the context of,
for example, manganites, where it is the ``Kondo'' coupling that 
is ferromagnetic due to Hund's rule.

\begin{figure}[htbp]
   \centering
   \includegraphics[width=3in]{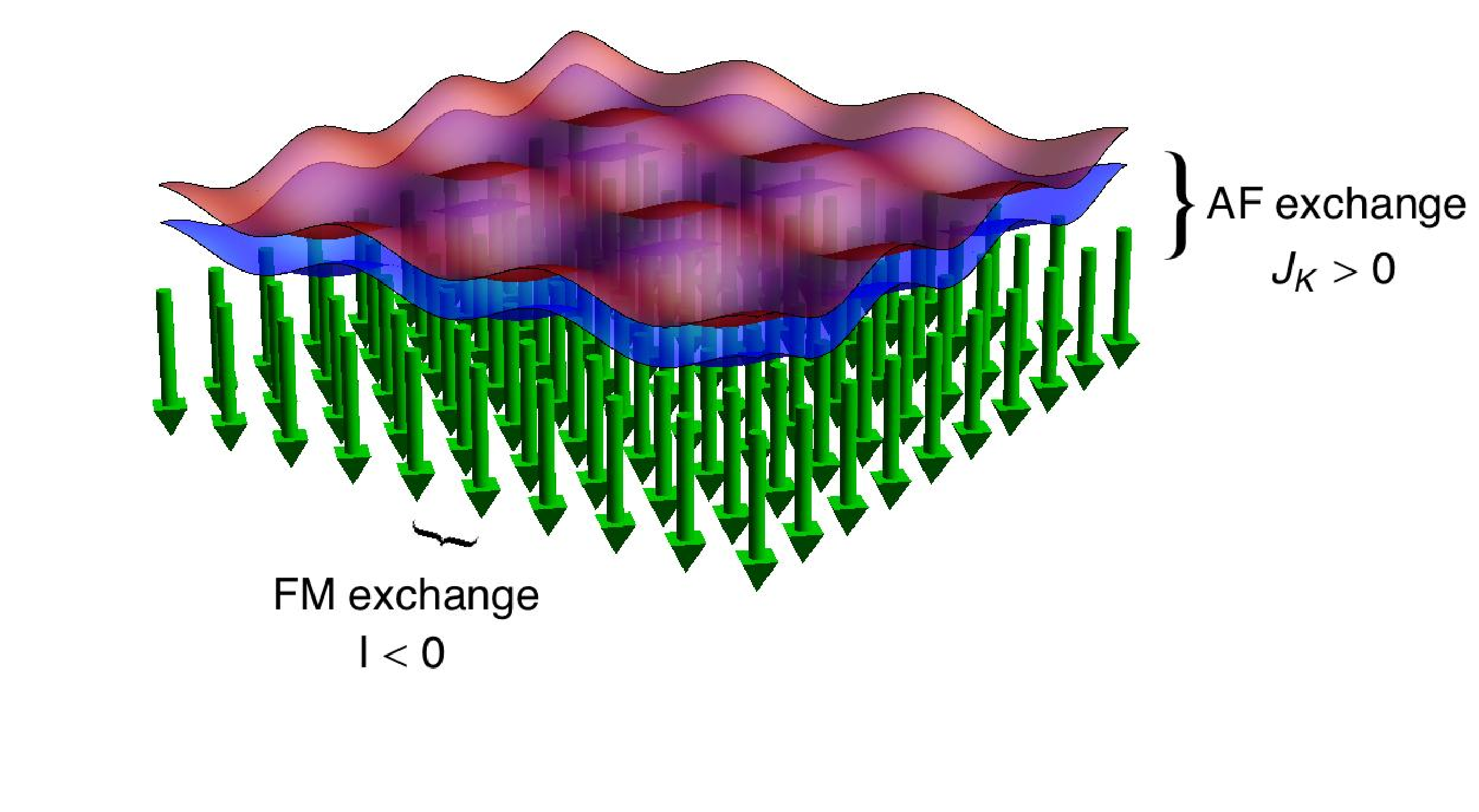}
   \caption{
   An illustration of the
   Kondo lattice. Local moments from f-orbitals are in green, and are
   depicted here to be spin down.
   Spin-up conduction electrons are in red, which have a higher
   probability density than the spin-down conduction electrons in blue.
   The Hamiltonian for the model is given in Eq.~(\ref{eq:KLHamiltonian})
   where $\sigma$ is the spin index and $a$ refers to the three
   spin directions.
   Note that the Einstein summation convention
   is used on indices.
   For simplicity, we assume $\epsilon_{\vec{K}} = \frac{K^2}{2m_e}$.
     The characteristic kinetic energy, $W$, is 
     defined as $W \equiv 1/\rho_0$, where 
     $\rho_0 \equiv \sum_{\vec{K}}\delta(E_F - 
     \epsilon_{\vec{K}})$ is the single-particle 
     density of states at the Fermi energy ($E_F$).
     Both $E_F$ and the chemical potential, $\mu$, scale 
     like $W$. We use the Shankar notation with 
     $K = |\vec{K}|$ measured from the center of the Brillouin zone.
}
   \label{fig:two_components}
\end{figure}

The parameter region we will focus on is $J_K \ll |I| \ll W$. Here 
we can use the limit $J_K=0$ as the reference point,
which contains the local moments,
representing the f-electrons with strong repulsions,
and conduction electrons.
As illustrated in Fig.~\ref{fig:two_components}, the local moments order in a ferromagnetic
ground state because $I<0$, whereas the conduction electrons form 
a Fermi sea with a Fermi surface. A finite but small $J_K$ will couple 
these two components,
and its effect is analyzed in terms of a fermion$+$boson 
renormalization group (RG) 
procedure~~\cite{Yamamoto07,Altshuler94,Shankar90}.
We will use 
an effective field theory approach, which we outline below and describe
in detail in the 
Supporting
Information.
Though our analysis will focus on this
weak $J_K$ regime,
the results will be germane to 
a more extended parameter
regime through continuity.

The Heisenberg part of the Hamiltonian,
describing the local moments alone,
is mapped to a 
continuum 
field theory~\cite{Read95}
in the form of a Quantum Nonlinear
Sigma Model (QNL$\sigma$M).
In this framework, the local moments are 
represented by an O(3) field, $\vec{m}$, which is constrained
non-linearly
with a continuum partition function.
Combining the local moments with the conduction electrons,
we reach the total partition function:
$Z= \int \mathcal{D}\vec{m}\mathcal{D}[\bar{\psi},\psi] \; 
\delta(\vec{m}^2(\vec{x},\tau)-1)e^{-\mathcal{S} }$,
where $\mathcal{S}=\mathcal{S}_{m}+\mathcal{S}_c^{\prime}+\mathcal{S}_K$.
The action for the conduction electrons,
$\mathcal{S}_{c}^{\prime}$,
is standard.
Defining $m^+ = m_x+im_y$ and $m^- = m_x - im_y$,
the low energy action for the local moments is
expressed in terms of a single complex scalar:
\begin{eqnarray}
\label{action-m}
	\mathcal{S}_m 
		&\approx& 
		\frac{1}{2}\int d\omega d^dq\; 
		m^{+}(\vec{q},i\omega) 
		(- M_0 i\omega + \rho_s q^2) 
		m^{-}(-\vec{q},-i\omega) \nonumber \\
		&&+ g\int \left(\partial m\right)^4
\end{eqnarray}
Here, $M_0$ is the magnetization density,
and $\rho_s$ the magnon stiffness constant.
The magnon-magnon coupling $g$, schematically written above
and more precisely specified in the 
Supporting
Information,
turns out to be irrelevant in the RG sense 
when fermions are also coupled to the system.
This 4-boson term, involving four gradients, was found to be 
relevant for $d>2$
in a model without fermions~\cite{Read95} but is unimportant when fermions are part of the system, as in our model here.
Finally, the Kondo coupling 
can be separated into static and dynamic parts.
The static order
of the local moments induces a 
splitting of the conduction electron
band on the order of $\Delta \sim J_K^z \langle m^z \rangle \sim J_K^z$,
which modifies $\mathcal{S}_c^{\prime}$ into 
the following action for the conduction electrons
\begin{eqnarray}\label{action-c}
	\mathcal{S}_{c}
	=
	\int d^dK d\epsilon\;\bar{\psi}_{\sigma}(\vec{K},\epsilon)
\left(-i\epsilon +\frac{K^2}{2m_e} - \mu + \sigma\Delta\right)
{\psi}_{\sigma}(\vec{K},\epsilon) \nonumber \\
\end{eqnarray}
The dynamical part couples the magnons with the conduction electrons,
leading to
\begin{eqnarray}
	\mathcal{S}_K^{\pm} &=& J_K^{\pm}\int d^d q d\omega d^dK d\varepsilon 
			\Big( 
			 \psi_{K+q ,\uparrow}^{\dagger} \psi_{K,\downarrow}^{\phantom\dagger}m^{-}_{q} \nonumber \\
			 &&+\quad
			 \psi_{K+q,\downarrow}^{\dagger}\psi_{K,\uparrow}^{\phantom\dagger}m^{+}_{q}
			\Big) \\
	\mathcal{S}_K^{z} &=& -\frac{J_K^{z} }{2} \int 
			d^d q_1 d\omega_1 d^d q_2 d\omega_2 d^dK d\varepsilon \nonumber \\
			&&\times \Big( 
			 \psi_{K+q_1-q_2 ,\sigma}^{\dagger} 
			 \tau^z_{\sigma \sigma '}\psi_{K,\sigma '}
			 ^{\phantom\dagger}
			 m^{-}_{q_1} m^{+}_{q_2}
			\Big)
			\label{action-Kondo}
\end{eqnarray}
The mapping from the microscopic model in 
Eq.~(\ref{eq:KLHamiltonian}) to the field 
theory in 
(\ref{action-m})-(\ref{action-Kondo})
is similar to the antiferromagnetic case~\cite{Yamamoto07},
but
differs from the latter in several important ways.
One simplification is that 
the translational
symmetry is preserved
in the ferromagnetic phase.
At the same time, two complications arise.
Ferromagnetic order breaks time-reversal symmetry,
which is manifested in the 
Zeeman splitting of the 
spin up and down bands.
In addition, the effective field theory for a local-moment 
quantum ferromagnet involves a Berry phase
term~\cite{Read95} such that Lorentz invariance is broken,
even in the continuum limit;
the dynamic exponent, connecting $\omega$ and $q$
in Eq.~(\ref{action-m}),
is $z=2$ instead of $1$.
The effective field theory, comprising 
Eqs.~(\ref{action-m})-(\ref{action-Kondo}),
is subjected to a two-stage RG analysis
as detailed in the 
Supporting
Information.

\begin{figure}[htbp]
   \centering
   \includegraphics[width=3in]{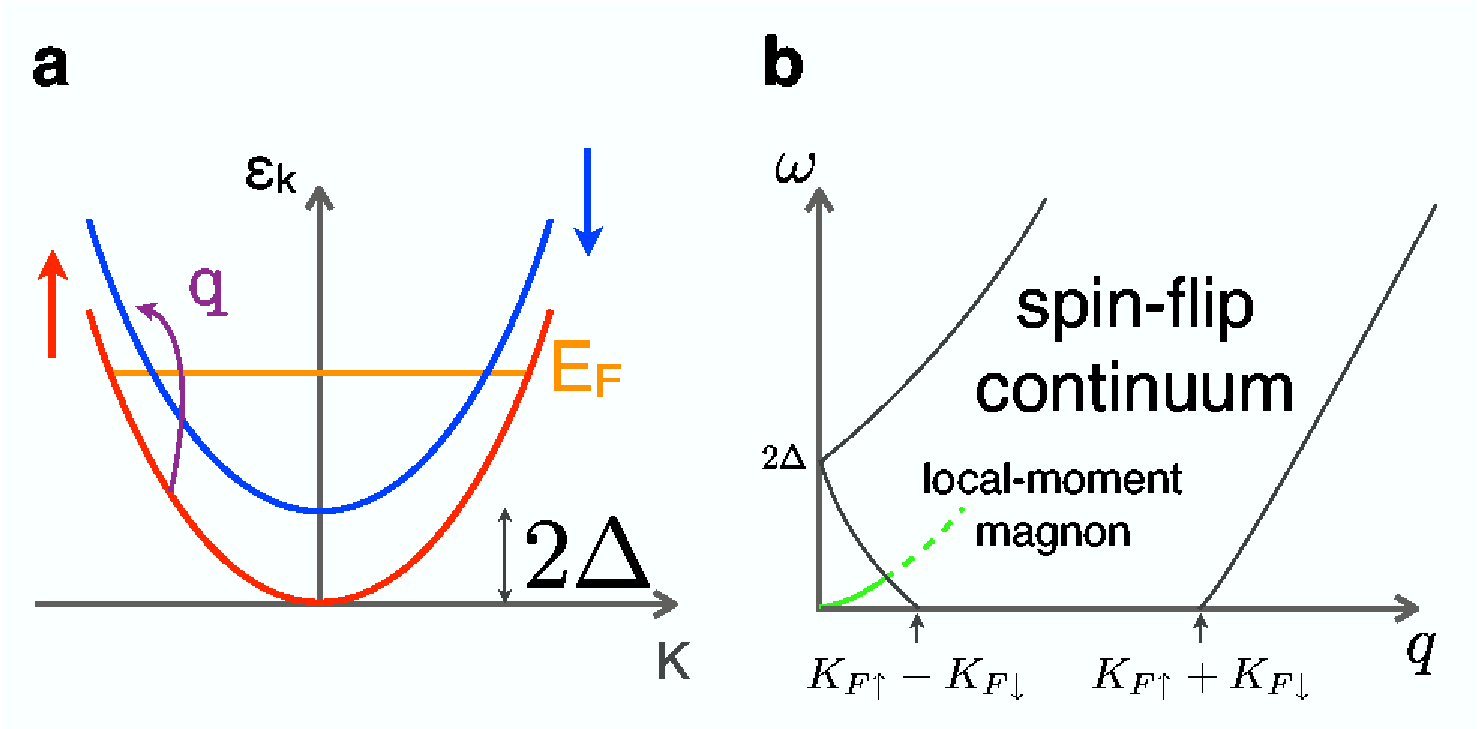}
   \caption{
   Phase space for the Kondo coupling. 
     {\bf a}, The spin-splitting of the conduction electron band, which
     kinematically suppresses interband processes 
     associated with the Kondo spin-flip coupling to
     the local-moment magnons.
     {\bf b}, The kinematics for the spin-flip Kondo coupling.
     The low-lying excitations
     of the local-moment system are the magnons
     which enter the continuum at finite $\omega$ and $q$. 
     Those of the conduction electrons
     are expressed in terms of the spin-flip continuum, whose Kondo-coupling
     to the local-moment magnons is 
     cut off below the cutoff energy, $\omega_c \approx (I/W^2)\Delta^2$,
     and the cutoff momentum, $q_c \approx K_{F\uparrow}-K_{F\downarrow}
     \approx (K_F/W)\Delta$.
}
   \label{fig:damping}
\end{figure}

\section{Results}

For energies and momenta
above their respective cutoffs,
$\omega_c \sim (I/W^2) \Delta^2$ 
and 
$q_c \sim
(K_F/W) \Delta$,
the magnons are coupled to the continuum part of the transverse 
spin excitations of the conduction electrons, 
see Fig.~\ref{fig:damping}.
Here, the Kondo coupling is relevant in the 
RG sense below three dimensions.
This implies strong coupling between the conduction electrons and the 
local moments, and both the QNL$\sigma$M as well as the
action for the conduction electrons will be modified. 
Explicitly, the correction to the quadratic part of the QNL$\sigma$M is 
\begin{eqnarray}\label{boson-self-energy}
	\Pi(\vec{q},\omega) 
		\approx J_K^2 \rho_0 \left( 1
+ i\gamma\frac{\omega}{v_F q} \right)
\end{eqnarray}
where $\gamma$ is a dimensionless constant prefactor.
At the same time, the conduction electrons
acquire the following self-energy:
\begin{eqnarray}\label{fermion-self-energy}
	\Sigma(K_F,\epsilon) &=& \left\{
		\begin{array}{cc}
			-A_2 (\rho_0 J_K^4/I^2)^{1/3} ~(-i\epsilon)^{2/3}
& ~~d=2 \\
-A_3(\rho_0J_K^2/I) ~\epsilon \log(-i\epsilon) 
& ~~d=3
		\end{array}\right.
		\label{eq:NFL}
\end{eqnarray}
where $A_2$ and $A_3$ are dimensionless constants of order unity.
The self energies $m^{+} \Pi m^{-}$ and $\bar{\psi} \Sigma \psi $  add directly 
to the quadratic parts of the action, $\mathcal{S}_m$ and $\mathcal{S}_c$, respectively.
Similar forms for the self-energies appear in other contexts,
notably the gauge-fermion
problem
and the spin-fluctuation-based quantum critical
regime.
The formal similarities as well as some of the important
differences are discussed in the 
Supporting
Information.

With these damping corrections
incorporated, the effective transverse Kondo coupling, $J_K^{\pm}$,
becomes marginal in the RG sense in both two and three dimensions;
the marginality is exact in the sense that it extends to infinite 
loops, as detailed in the 
Supporting
Information.
This signals the stability of the form of damping for
both the magnons and conduction electrons~\cite{Altshuler94,Polchinski93}.
At the same time, the effective longitudinal Kondo coupling,
$J_K^{z}$, as well as the non-linear coupling among the magnons,
$g$, are irrelevant in the RG sense.

The exactly marginal nature of the Kondo coupling in the continuum part 
of the phase 
space implies that the effective coupling remains small as we scale down to
the energy cutoff 
$\omega \sim \omega_c$ and, correspondingly,
the momentum cutoff $q \sim q_c$.
Below these cutoffs, the transverse Kondo coupling, which
involves spin flips of the conduction electrons, cannot 
connect two points near the up-spin and down-spin Fermi surfaces;
see Fig.~\ref{fig:damping}.
Although there is no gap in the density of states, 
as far as the spin-flip Kondo coupling is concerned,
the system behaves
as if the lowest energy excitations have been gapped out.
The important conclusion, then, is that the effective transverse 
Kondo coupling renormalizes to zero 
in the zero-energy and zero-momentum limit.
This establishes the absence of static Kondo screening.
Hence, the Fermi surface is small, and this is illustrated in
Fig.~\ref{fig:small_large_FS}a.

The region of validity of 
Eqs.~(\ref{boson-self-energy},\ref{fermion-self-energy})
corresponds to $\omega_c \ll \omega \ll |I|$ 
and $q_c \ll q \ll 2K_F$. 
This range is well-defined,
given that $\Delta \approx J_K \langle m^z \rangle \leq J_K$
and that we are considering
$J_K \ll |I| \ll W$.
In this same energy and, correspondingly, temperature ranges,
other physical properties also show a non-Fermi liquid behavior.
In two dimensions, the specific heat coefficient, $C/T \sim T^{-1/3}$
and the electrical resistivity $\rho \sim T^{4/3}$.
In three dimensions, $C/T \sim \log(1/T)$ and 
$\rho \sim T^{5/3}$. These non-Fermi liquid features have 
form similar to those of the quantum critical 
ferromagnets~\cite{Smith_Nature08,Belitz_RMP05}, 
although here we are deep
inside the ferromagnetically-ordered part of the phase diagram.

\section{Discussion}

Our result is surprising given that the ratio $J_K/\omega_c 
\sim W^2/\left(I J_K\langle m^z \rangle ^2\right) \gg 1$.
By contrast, the standard Kondo impurity problem 
with a pseudo-gap of order $\Delta_{pg} \ll J_K$ 
in the conduction electron density 
of states near the Fermi energy would be
Kondo-screened~\cite{Ingersent98, Withoff90}.
The difference is that, in the latter case, the Kondo coupling renormalizes
to stronger values as the energy is lowered in the range $\Delta_{pg} \ll
\omega \ll W$; for $J_K/\Delta_{pg} \gg 1$, 
the renormalized Kondo coupling 
is already large by the time the energy
is lowered to $\omega \sim \Delta_{pg}$.

\begin{figure}[htbp]
   \centering
   \includegraphics[width=3in]{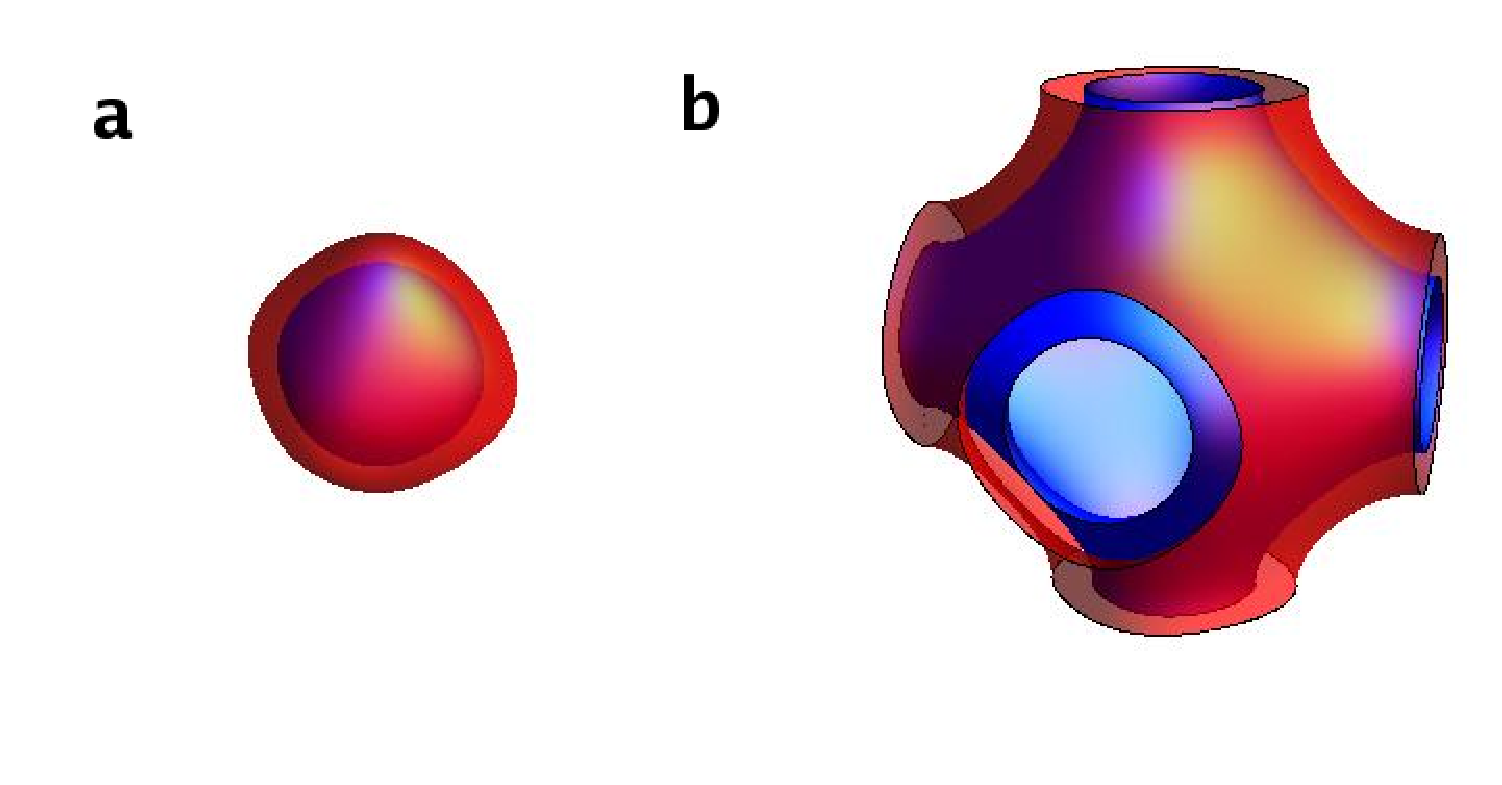}
   \caption{
   Contrasting the small and large Fermi surfaces.
   The spin-up electron Fermi surface is drawn in red
   and larger
   than the spin-down electron Fermi surface in blue.
   The larger Fermi surface has been made slightly transparent
   to reveal the smaller sheet.
   {\bf a}, The local moments are not part of the Fermi surface.
   {\bf b}, The static Kondo screening has caused the
   Fermi surface to expand to accommodate the Kondo resonances 
   associated with the local moments.
   }
   \label{fig:small_large_FS}
\end{figure}

The small Fermi surface we have established 
is to be contrasted with the large Fermi surface
of a ferromagnetic heavy fermion metal in the Stoner treatment,
illustrated in Fig.~~\ref{fig:small_large_FS}b.
In the latter case, 
the local moments become entangled with the conduction electrons
as a result of the static Kondo screening.
Kondo resonances develop
and the local moments become incorporated
into a large Fermi surface.
This Fermi surface 
comes from a Zeeman-splitting of an
underlying Fermi surface for the paramagnetic phase;
the latter is large, as seen through 
a non-perturbative proof~\cite{Oshikawa00} that 
relies upon
time-reversal invariance.

Our result of a stable ferromagnetic metal phase
with a small Fermi surface provides the basis to understand
the dHvA-measured~\cite{King91, Yamagami94, Ikezawa97}
Fermi surface of CeRu$_{\rm 2}$Ge$_{\rm 2}$,
which is ferromagnetic below $T_c = 8$ K.
Our interpretation rests on a dynamical Kondo screening effect that
turns increasingly weak at lower energies. This is supported by
the observation of the collapsing quasielastic peak measured in 
the inelastic neutron-scattering cross section as the temperature
is reduced~\cite{Rainford96}.
It will be very instructive if 
the Fermi surface of UGe$_2$~\cite{Settai02} is further clarified
and 
if systematic dHvA measurements 
are carried out in 
other ferromagnetic 
heavy fermion metals as well.
With future experiments in mind, 
we note that our conclusion of a small Fermi surface also
applies to ferrimagnetic order.

In the parameter regime we have considered, the non-Fermi liquid features
are sizable. For instance, the non-Fermi liquid contribution to the 
self-energy [Eq.~(\ref{fermion-self-energy})] is,
at the cutoff energy $\omega_c$, larger than the standard Fermi liquid
term associated with the interactions among the conduction electrons.
It remains to be fully established whether the non-Fermi liquid
terms in the electrical resistivity and specific heat
can be readily isolated from contributions of other processes.
Still, there is at least one family of materials,
URu$_{\rm 2-x}$Re$_{\rm x}$Si$_{\rm 2}$ at 
${\rm x}>0.15$,
in which non-Fermi liquid features have been shown to
persist deep inside the ferromagnetic regime~\cite{Bauer05,Butch09}.
Whether this observed feature
is indeed a property of the ferromagnetic phase, 
or if it is related to some quantum critical fluctuations or
even certain disorder effects,
remains to be clarified 
experimentally. 
We hope that our theory will provide
motivation 
for the experimental search of non-Fermi liquid behavior in
ferromagnetic heavy fermion metals as well.

The existence of a ferromagnetic phase with a small Fermi surface
raises the prospect of a direct quantum phase transition 
from a Kondo-destroyed ferromagnetic metal to a Kondo-screened 
paramagnetic metal.
This, like its 
antiferromagnetic
counterpart~\cite{Gegenwart08,Paschen04,Park08}, in turn raises the
possibility of a new type of superconductivity;
the underlying quantum fluctuations  
would be associated with not only the development of the ferromagnetic
order~\cite{Saxena00}
but also the transformation of a large-to-small Fermi surface.
Theoretically, accessing the quantum phase transition requires
that our analysis be extended to the regime where the Kondo coupling
is large compared to the RKKY interaction, and this represents an
important direction for the future. Experimentally, 
in the case of CeRu$_{\rm 2}$Ge$_{\rm 2}$, applying 
pressure or doping Ge with Si fails to reach a 
ferromagnetic-to-paramagnetic quantum phase transition
due to the intervention of antiferromagnetic order; other 
control parameters or other ferromagnetic heavy fermions
should be explored.

Finally, it is instructive to compare our study of the Kondo lattice 
Hamiltonian with traditional studies of itinerant ferromagnetism based
on a one-band Hubbard model. We have taken advantage of the separation
of energy scales of the Kondo lattice Hamiltonian and derived our results
from an asymptotically exact RG analysis. By contrast, the one-band Hubbard
model does not feature a separation of energy scales at the Hamiltonian
level and the corresponding theoretical studies ~\cite{Ueda75} have been
based on mean-field (randon-phase) approximations. Furthermore,
the separation of energy scales in the Kondo lattice Hamiltonian 
is crucial for our conclusion that the non-Fermi liquid behavior
exists over a large energy window, which does not happen in the one-band
case. Finally, the issue of small Fermi surface, which represents a major
conclusion of our study, is absent in the case of the one-band Hubbard
model.

To summarize, we have shown that the ferromagnetic Kondo lattice 
has a parameter range where the Kondo screening is destroyed and 
the Fermi surface is small. 
This conclusion is important for heavy fermion physics.
It allows us to understand a long-standing puzzle on the Fermi surface,
as epitomized by the dHvA measurements in CeRu$_2$Ge$_2$.
It also sharpens the analogy with the extensively studied
antiferromagnetic
heavy fermion metals, where the dichotomy between Kondo breakdown
and conventional quantum criticality is well established.
More broadly, the present work has led to 
one of the very few asymptotically
exact results for metallic ferromagnetism
whose rigorous understanding has remained elusive for many years~\cite{Vollhardt98}. 
Our findings highlight an important lesson,
namely that correlation effects can lead to qualitatively
new properties even for magnetism occurring in a metallic environment.  
This general lesson 
could very well be relevant to a broad array of magnetic 
systems, including 
the extensively-debated iron pnictides~\cite{Cruz08}.

\begin{acknowledgments}
We acknowledge
M. C. Aronson, N. P. Butch, S. R. Julian, Y. B. Kim, 
P. Goswami, M. B. Maple, C. P{\'e}pin, T. Senthil, 
and I. Vekhter for discussions,
and the National Science Foundation
and the Robert A. Welch Foundation Grant C-1411 for partial support.
\end{acknowledgments}

%% file: SI.tex



\begin{center}
{\large\bf 
Supporting Information for: \\
``Metallic Ferromagnetism in the Kondo Lattice''
}
\\[0.5cm]

Seiji J. Yamamoto\\
{\em NHMFL and FSU Department of Physics, Tallahassee, FL 32310, USA}
\\[0.5cm]
and
\\[0.5cm]

Qimiao Si \\
{\em Department of Physics and Astronomy, Rice University, Houston, TX 77005, USA}\\

\end{center}

\section{Kondo Lattice and Field Theory}
We begin with a microscopic description of heavy fermion metals
in terms of the Kondo-Heisenberg Hamiltonian.
\begin{eqnarray}\label{eq:KLHamiltonianSupp}
	H
	&=&
	\sum_{k} \epsilon^{\phantom\dagger}_{\vec{K}} c^{\dagger}_{\vec{K}\sigma}c^{\phantom\dagger}_{\vec{K}\sigma} + 
	I \sum_{\langle ij \rangle} S^a_i S^a_j  +
		\sum_{i}
J_K^a S^a_i c^{\dagger}_{i\sigma} \frac{\tau^a_{\sigma\sigma^{\prime} } }{2}c^{\phantom\dagger}_{i\sigma^{\prime} } \nonumber \\
\end{eqnarray}
where $a$ labels the three spin components.
For simplicity, and without loss of generality,
we will consider only nearest-neighbor ($\langle ij \rangle$)
ferromagnetic exchange interaction among the local moments,
and we will also assume $\epsilon_{\vec{K}} = \frac{K^2}{2m_e}$. 
By contrast to the purely itinerant magnets, the local moments are 
independent degrees of freedom to begin with and, on their own,
would be
ferromagnetically ordered
($I<0$).
These local moments are also antiferromagnetically
coupled ($J_K>0$) to itinerant conduction 
electrons. 
The exchange interaction among the local moments includes 
not only 
the RKKY
interaction generated by the conduction electrons in 
Eq.~(\ref{eq:KLHamiltonianSupp}), but also the RKKY and superexchange
interactions from other conduction-electron states as well as 
the direct exchange. We note that our model is very
different from those for double-exchange ferromagnets in the context
of the manganites, where the ``Kondo'' coupling 
is ferromagnetic~\cite{Pandey08_SI,Kapetanakis08_SI}:
there, the low-lying states are spin triplets formed by the local spins and 
itinerant conduction electrons, and the Kondo screening physics 
is never pertinent.

Since we are interested in the low energy properties of the ferromagnetic
phase of this system, we adapt an effective field theory previously used 
for the pure quantum Heisenberg ferromagnet~\cite{Read95_SI},
but extend it to include fermions.  Here, the spin is represented by 
an O(3) field, $\vec{m}$, which is constrained non-linearly.
\begin{eqnarray}\label{eq:KLFieldTheory1}
	Z &=& \int \mathcal{D}\vec{m}\mathcal{D}[\bar{\psi},\psi] \; \delta(\vec{m}^2(\vec{x},\tau)-1)e^{-\mathcal{S} } \nonumber \\
	\mathcal{S} &\equiv& \mathcal{S}_{m}^{\prime}+\mathcal{S}_{\text{Berry}}+\mathcal{S}_{c}^{\prime}+\mathcal{S}_{K} \nonumber\\
	\mathcal{S}_{m}^{\prime} &=& \frac{\rho_s}{2}\int d^dx d\tau\; \frac{\partial m^a(\vec{x},\tau)}{\partial x^{\mu}}\frac{\partial m^a(\vec{x},\tau)}{\partial x^{\mu}} \nonumber \\
	\mathcal{S}_{\text{Berry}} &=& iM_0 \int d^dx d\tau\; A^a[\vec{m}] \frac{\partial m^a(\vec{x},\tau)}{\partial \tau} \nonumber \\
	\mathcal{S}_{c}^{\prime} &=& \int d^dx d\tau\; \bar{\psi}_{\sigma}(\vec{x},\tau)\left(\partial_{\tau} -\frac{\nabla^2}{2m_e}-\mu\right)\psi_{\sigma}(\vec{x},\tau)  \nonumber \\
	\mathcal{S}_K &=& J_K^{a} \int d^d x d\tau \; s_c^a(\vec{x},\tau) m^a(\vec{x},\tau)
\end{eqnarray}
where, as usual, $s_c^a \equiv \bar{\psi}_{i\sigma} \frac{\tau^a_{\sigma\sigma^{\prime} }
}{2}\psi_{i\sigma^{\prime} }$, and the superscript $\mu$ labels the spatial directions.  
The topological Berry phase term is crucial to get
the dynamics right~\cite{Wen88_SI}.
If we define the 
$z$-axis as the direction of magnetization, we have $\nabla_m \times 
\vec{A} = (0,0,1) = \langle \vec{m} \rangle$ (note that the curl is 
in field space, not real space).  Thus, in a linearized, low-energy 
theory of spin fluctuations, we have $\vec{A} \approx (-m_y, m_x,0)$.
Defining $m^+ = m_x+im_y$ and $m^- = m_x - im_y$ we obtain a theory 
of a single complex scalar
\begin{eqnarray}\label{eq:KLFieldTheory2Supp}
	\mathcal{S}_m 
		&=& \mathcal{S}_{m}^{\prime} +  \mathcal{S}_{\text{Berry}} \\
		&\approx& \frac{1}{2}\int d\omega d^dq\; 
		m^{+}(\vec{q},i\omega) 
		\left(- M_0 i\omega + \rho_s q^2 \right) 
		m^{-}(-\vec{q},-i\omega) \nonumber
\end{eqnarray}
We have now arrived at an effective theory of local moment ferromagnetic 
magnons
coupled to fermions with effective coupling constant that for simplicity we also label $J_K$.
The mapping from the microscopic model in equation 
(\ref{eq:KLHamiltonianSupp}) to the field theory
in (\ref{eq:KLFieldTheory1})
parallels the antiferromagnetic (AF) 
case~\cite{Yamamoto07_SI, Read95_SI}.

\section{Scaling Analysis}
We need to carry out an RG analysis for the field
theory above several times, both before
and after self-energies have been incorporated.
To begin, we summarize the pure boson problem which has been done previously~\cite{Read95_SI}.
The dimension of the $m$ field is fixed by the nonlinear constraint
$m^a(\vec{x},\tau)m^a(\vec{x},\tau) =1$ which requires $[m^a(\vec{x},\tau)]=0$. 
In momentum space, this becomes $[m^a(\vec{q},\omega)]=-d-z_b$.
Unless indicated otherwise, we will exclusively be concerned
with field dimensions in momentum space, 
so the arguments will often be dropped: $[m] = -d-z_b$.
As usual for purely bosonic RG, the momenta
and energies scale simply as $[q]=1$ and $[\omega]=z_b$,
where $z_b = 2$ is the dynamical exponent for the boson, 
which is consistent with $\omega \sim q^2$.  
The modulo $4\pi$ ambiguity in the Berry phase
dictates $[M_0] = d$, and the scale invariance of
$\mathcal{S}_m$ establishes $[\rho_s]=d+z_b-2$.

Read and Sachdev were the first to point out that higher order gradient
terms may be relevant.
\begin{eqnarray}\label{non-linear}
	\mathcal{S}_m^{(4)} &=& g \int d^dx d\tau 
	\Big( \partial_{\mu} m_a  \partial_{\mu} m_a  \partial_{\nu} m_b  \partial_{\nu} m_b \nonumber\\
	&&\quad-2 \partial_{\mu} m_a \partial_{\nu} m_a \partial_{\mu} m_b \partial_{\nu} m_b \Big)
\end{eqnarray}
Using the scaling scheme described above, this coupling, representing
magnon-magnon interactions, has scaling dimension $[g]=d-2$.
This indicates that, for $d > 2$, the magnon-magnon scattering 
is relevant.
We will see later why this term becomes irrelevant when fermions
are incorporated.

In parallel to the pure boson problem, there is a well known
procedure for handling pure fermion problems within a
momentum shell approach~\cite{Shankar94_SI}.
The essential
difference from the bosonic RG is that the low energy manifold
now consists of an extended surface, the Fermi surface,
rather than a single point.  Scaling should therefore be done
with respect to this surface, and this may be accomplished
by a clever change of coordinates for a simple spherical Fermi
surface.

When the action contains both bosons and fermions, the momentum shell
RG becomes much more complicated.  In the special case
$z_f=1$ and $z_b=1$, we have extended Shankar's approach
in a straightforward fashion~\cite{Yamamoto07_SI}.
However, such an approach does not work if $z_f \neq z_b$.
Another strategy has been proposed by Altshuler, Ioffe, and Millis~\cite{Altshuler94_SI},
and we adopt this method here.
For further details on this method, see~\cite{Yamamoto10_SI}

Each fermion momentum space integral is decomposed into patches of size 
$\Lambda_f$ in every direction so that each patch is locally
a flat space.  Scaling is accomplished locally with respect to the 
center of each patch.  Momenta are therefore decomposed
into components parallel ($k_{\parallel}$) and perpendicular ($k_{\perp}$)
to the vector normal
to the Fermi surface at this reference point.  For example,
$\int_{\text{annulus}} d^dK = \sum_{\text{patches}} 
\int_{-\Lambda_f}^{\Lambda_f} d^{d-1}k_{\perp}dk_{\parallel}$.  
Note that some authors use an opposite naming convention
for components; we follow the notation of Ref.~\cite{Altshuler94_SI}.
A tacit assumption of this approach is that
the boson does not connect two fermions in different
patches; this is only justified for forward scattering problems
like the one we consider in this paper.
Bosonic momentum integrals are already constrained to
a volume of linear dimension $\Lambda_b$, which we assume
naturally fits inside the fermionic patch: 
$\Lambda_b \sim \Lambda_f \equiv \Lambda$.
In this scheme, fermionic and bosonic momenta scale the same way,
albeit anisotropically.
The assignment of values for $[\epsilon]$, $ [k_{\parallel}] $, 
and $[k_{\perp}]$
will depend on the form of the quadratic action,
and this will
be different depending on how we incorporate the 
corrections to the QNL$\sigma$M and fermion actions.
The scaling analysis will
therefore need to be done anew for each case.

The introduction of fermions and the choice
to use the scaling procedure outlined above has
an immediate consequence on the way we scale the
bosonic action.
In the pure boson case, we can use $[M_0]=d$.
This comes from the modulo $4\pi$ ambiguity of the 
Berry phase.  Specifically, since $e^{i4\pi S} = 1$, we need
$i4\pi S = i 2\pi n$, where $n$ is an integer.  
Therefore $S$ is quantized at either an integer or
half integer value, and is insensitive to the RG rescaling.  
However, since $S = M_0 \int d^dx = M_0 L^d$, 
and since $[L^d] = -d$, we must have $[M_0] = d$.~\cite{Sachdev_book_SI}
But the anisotropic scalings
we employ in momentum space no longer translate
simply to a real space analysis.  We must therefore
abandon these dimension assignments for the pure boson
problem.  Instead, we write the action completely in momentum space
and live with the understanding that after rescaling, the fields
$m^a(\vec{q},\omega)$ and $\psi(\vec{k},\epsilon)$
no longer represent the Fourier transforms of
the microscopic fields 
$m^a(\vec{x},\tau)$ and $\psi(\vec{x},\tau)$.
This is nothing new since even in the original
Wilsonian RG formalism the imposition of a cutoff
invalidates the interpretation of $\phi(q)$ as a true
Fourier transform of $\phi(x)$.

A second reason to modify the Read-Sachdev assignments
for scaling dimensions in the pure boson problem is that the addition
of fermions acts as a magnetization sink for the local-moment system.
Of course, the overall magnetization is still conserved in the
ferromagnetic phase. 
Furthermore, we assume there are no
valence fluctuations (an implicit assumption in writing down 
the microscopic Kondo-Heisenberg Hamiltonian) so we can 
still treat the local moments as O(3) spins attached to the lattice,
and therefore work with the nonlinear field theory.

The way we fix the scaling dimensions is to define the quadratic action
according to:
\begin{eqnarray}
	\mathcal{S}_m 
		&=& \int d\omega\; d^{d-1}q_{\perp}dq_{\parallel}\; 
		m^{+}\left(-i\omega + q_{\perp}^2 \right)m^{-} \\
	\mathcal{S}_c
		&=& \sum_{\text{patches}} \int d\epsilon\; d^{d-1}k_{\perp}dk_{\parallel}\; 
		\bar{\psi}_{\sigma}
		\left(i\epsilon  - v_F k_{\parallel} - \frac{v_F}{2K_F}k_{\perp}^2 \right) 
		\psi_{\sigma} \nonumber \\
\label{eq:quadraticFermionPart}
\end{eqnarray}
where, as usual~\cite{Altshuler94_SI}, 
$q_{\perp} \gg q_{\parallel}$.
The coupling of the local-moment magnons to the fermions introduces 
anisotropy in momentum space; as we will see, such an anisotropic
fixed point turns out to be exactly marginal.
To ensure that these forms are scale invariant, we make the assignments:
\begin{eqnarray}
	\left[ \epsilon \right] &=& 1 \nonumber
\nonumber
\\
	\left[ k_{\parallel} \right] &=& 1
\nonumber
\\
	\left[ k_{\perp} \right] &=& 1/z_b = 1/2
\nonumber
\\
	\left[ \psi \right] &=& -(3z_b+d-1)/(2z_b) = -(5+d)/4
\nonumber
\\
	\left[ m \right] &=& -(2z_b+d+1)/(2z_b) = -(5+d)/4
\end{eqnarray}
This information is used to count dimensions for the Kondo coupling (see figure \ref{fig:vertices}).
\begin{eqnarray}
	\mathcal{S}_K^{\pm} &=& J_K^{\pm}
		\int d^{d-1} q_{\perp}dq_{\parallel} d\omega  d^{d-1} k_{\perp}dk_{\parallel} d\varepsilon \nonumber \\
&&\times		\Big[ 
			 \bar{\psi}_{k+q,\uparrow}\psi_{k,\downarrow}m^{-}_{q} +
			 \bar{\psi}_{k+q,\downarrow}\psi_{k,\uparrow}m^{+}_{q}
			\Big] 
\end{eqnarray}
The tree-level dimension of the Kondo coupling is now easily found.
\begin{eqnarray}
	[ \mathcal{S}_K^{\pm} ] &=& 0  \nonumber \\
		&=& [J_K^{\pm}] +2[d^{d-1}k_{\perp}dk_{\parallel} d\varepsilon] + 2[\psi] + [m] \nonumber \\
		&=& [J_K^{\pm}] + 2\frac{d-1+2z_b}{z_b}-2\frac{3z_b+d-1}{2z_b} \nonumber \\
		&&-\frac{2z_b+d+1}{2z_b} 
		\nonumber \\
\implies [J_K^{\pm}] &=& (3-d)/(2z_b)
\end{eqnarray}
The spin-flip Kondo coupling is relevant in
two dimensions, and marginal (at the tree level) in three dimensions.
Usually, when the Kondo coupling is relevant, 
we expect the model to flow
to a strong coupling fixed point where Kondo screening sets in, destroying the magnetic order and
leading to a paramagnetic phase with a large Fermi surface.  This, however, would be an 
incorrect, and inconsistent, conclusion.  
A proper calculation of the self energies and subsequent
re-analysis of the scaling dimensions around the
appropriate fixed point will show that there will
never be Kondo screening.

\section{Damping correction to the QNL$\sigma$M and Scaling}
Our analysis so far has been a little too naive.  In particular,
it describes the wrong fixed point.
Note that so far we have not considered the $z$-component
of the Kondo interaction, $J_K^z \int s_c^z m^z$, which we
refer to as the longitudinal channel.
This coupling has two important effects.  
First, it introduces the effect of splitting
the spin bands of the conduction electrons.
Second, 
when the modified bosonic propagator is inserted into the
fermionic self energy we will obtain a non-Fermi liquid form when
the Kondo coupling is SU(2) symmetric 
($J_K^+=J_K^{-}=J_K^z$).
What is crucial for this, of course, is that the 
magnons will remain gapless in the presence of the Kondo
coupling to the conduction electrons, and we wish to
show this explicitly.
With all this in mind, we present below in some detail the 
calculation of the magnon self-energy,
as well as an RG analysis with the modified 
QNL$\sigma$M.

\begin{figure*}
   \centering
   \includegraphics[width=3in]{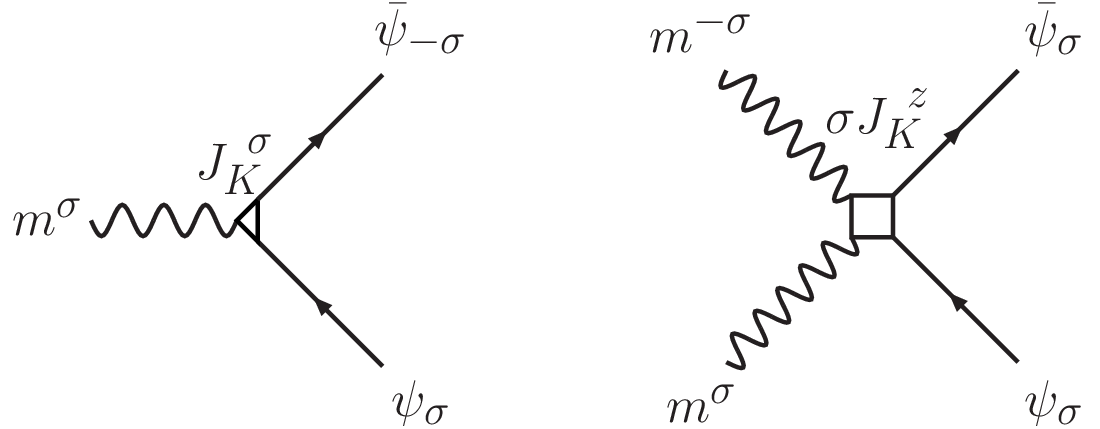}
   \caption{Interaction vertices}
   \label{fig:vertices}
\end{figure*}

\begin{figure*}
   \centering
   \includegraphics[width=3in]{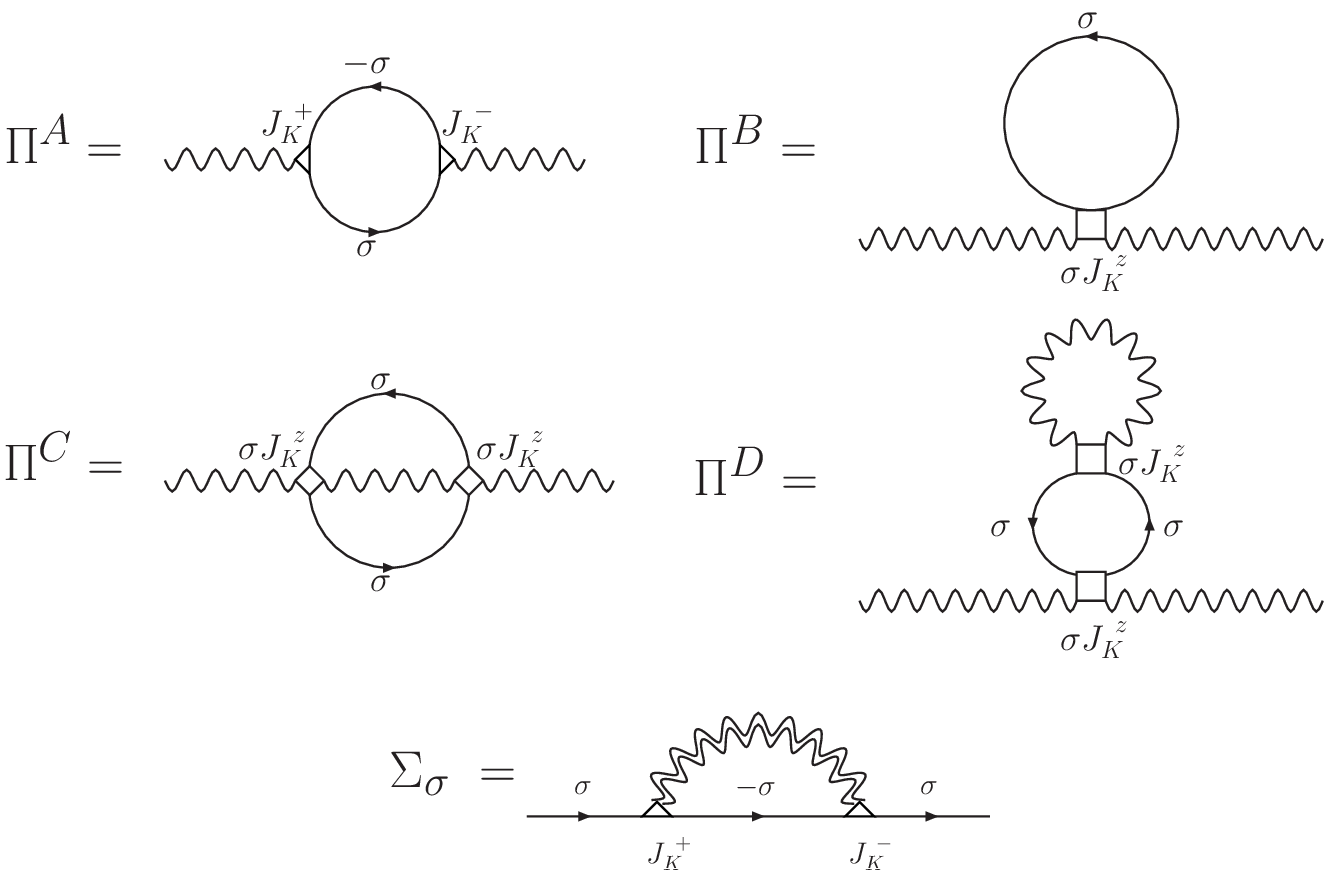}
   \caption{Self Energies}
   \label{fig:self-energies}
\end{figure*}

The first observation is easy to demonstrate.
For small fluctuations about the ordered state,
the longitudinal interaction is approximately 
$J_K^z \int (\bar{\psi}_{\uparrow}\psi_{\uparrow} - 
\bar{\psi}_{\downarrow}\psi_{\downarrow})(1-\frac{1}{2}m^{+}m^{-} )$.  
where we have used the constraint $m^z = \sqrt{1-m^{+}m^{-} }$.
The ``1'' comes from the 
magnetization in the $z$-direction, and leads to a Zeeman
shift in the energy of the conduction electrons.
The reference point for our
theory should therefore have
a quadratic action for the fermions of the form
\begin{eqnarray}
	\mathcal{S}_{c}
	&=& 
	\int d^dx d\tau\;\bar{\psi}_{\sigma}(\vec{x},\tau)\left(\partial_{\tau} -\frac{\nabla^2}{2m}-\mu + \sigma\Delta\right)\psi_{\sigma}(\vec{x},\tau)  \nonumber \\
\end{eqnarray}
where $\Delta \sim J_K^z \langle m^z \rangle \sim J_K^z$.
We need to write this in momentum space where it has the effect of defining
a spin-dependent Fermi wavevector: $K_{F\sigma} 
\equiv \sqrt{2m_e(\mu+\sigma\Delta)} $.  Expression 
(\ref{eq:quadraticFermionPart}) is unchanged except for the new definition
of $K_{F\sigma}$.
We need to build an effective low-energy theory around this fixed point, 
where there is a gap of size $2\Delta$
between the up-spin and down-spin bands.
This form of the fermionic spectrum is essential to correctly 
capture the damping of magnons via the Kondo interaction.  
The interaction vertices are represented diagrammatically 
in figure \ref{fig:vertices}, while the leading contributions
to the self energies are shown in figure \ref{fig:self-energies}.
The real and imaginary parts of the retarded functions can be
calculated exactly.  For example, the contribution from diagram
$\Pi^A$ is
\begin{eqnarray}
	\Re \Pi_R^{A}(\vec{q},\omega) 
		&=& -\frac{mJ_K^+J_K^{-}}{ q\pi } \Big[
			q  \nonumber \\
&&			+K_{F\uparrow}\text{sgn}(\zeta_{-,\uparrow}) \Theta(|\zeta_{-,\uparrow}|-1)\sqrt{\zeta_{-,\uparrow}^2 - 1} \nonumber \\
&&		   +K_{F\downarrow} \text{sgn}(\zeta_{+,\downarrow}) \Theta(|\zeta_{+,\downarrow}|-1)\sqrt{\zeta_{+,\downarrow}^2 - 1} 
			 \Big] \nonumber \\
	\Im \Pi_R^{A}(\vec{q},\omega) 
		&=& \frac{mJ_K^+J_K^{-}}{q\pi} 
\Big[ 	-K_{F\uparrow} \Theta(1-|\zeta_{-,\uparrow}|)
\sqrt{1-\zeta_{-,\uparrow}^2} \nonumber\\
&&+K_{F\downarrow} \Theta(1-|\zeta_{+,\downarrow} |)
\sqrt{1-\zeta_{+,\downarrow}^2} 			
\Big]			
\end{eqnarray}
where we have defined $\zeta_{\pm,\sigma} \equiv \frac{\omega - 2\Delta}{v_{F\sigma}q} \pm \frac{q}{2K_{F\sigma}} $, and $\sigma \in \{+,-\}$.  
The region in $(\omega,q)$-space where the imaginary part is non-zero 
is depicted in the main paper.  A similar exact expression is also 
available in $d=3$, but the approximate form is perhaps more useful.
The bubble $\Pi_R^A$ in the regime 
$\Delta \ll \omega \ll v_F q \ll 
\mu = K_F^2/(2m_e)$ is approximately:
\begin{eqnarray}
	\Pi_R^{A}(\vec{q},\omega) 
		&\approx& J_K^+J_K^{-} \rho_0^{(d)} \left( 
	1 + i \gamma_d\frac{\omega}{v_F q} \right)
\label{qnlsm-damping}
\end{eqnarray}
where $\gamma_d$ is a constant prefactor 
which depends on the spatial dimension, 
and $\rho_0^{(d)} =  \sum_{\sigma}\rho_{0,\sigma}^{(d)}$ is the density
of states at the Fermi level.  In two and three dimensions, the explicit
expressions are $\rho_{0,\sigma}^{(d=2)} = \frac{m}{2\pi}$ and 
$\rho_{0,\sigma}^{(d=3)} = \frac{m}{2\pi^2}K_{F\sigma}$.  
The $\omega/q$ form of the damping is common to a variety of systems;
in this case it signifies Landau damping of the magnons
with spin 1 excitations of the fermions.

To satisfy Goldstone's theorem,
it is necessary for all the pieces of $\Pi$ to cancel in such a way that 
the full bosonic propagator emerges in massless form.  In the gauge-fermion
problem, this is a consequence of gauge invariance~\cite{Tsvelik_book_SI}.
In our case,
the cancellation is somewhat more subtle.  First, note that the diagrams $\Pi^C$ and $\Pi^D$ are explicitly $O(J_K^2)$.  Diagrams $\Pi^A$ and $\Pi^B$, however, are both linear in $J_K$.  This is obvious for $\Pi^B$, whose calculation is trivial:
\begin{eqnarray}
	\Re \Pi_R^{B}(\vec{q},\omega) 
		&=& -J_K^z(n_{\uparrow} - n_{\downarrow}) \nonumber \\
	\Im \Pi_R^{B}(\vec{q},\omega) 
		&=& 0
\end{eqnarray}
The sign difference comes from the fact that there is a four-leg vertex $J_K^z$ for each spin, but the sign of the coupling constant depends on $\sigma$.  The reason why $\Pi^A$ is linear in $J_K$ instead of $O(J_K^2)$ can be seen from a simple calculation at $(\vec{q}=0, \omega=0)$, which is non-singular due to the different spin indices.  After performing the Matsubara sum,
\begin{eqnarray}
	\Pi^A_R(\vec{0},0) 
	&=& 2J_K^{+}J_K^{-} \int \frac{d^dK}{(2\pi)^d}\frac{n(\xi_{K,\uparrow}) - n(\xi_{K,\downarrow} ) }{\xi_{K,\downarrow} - \xi_{K,\uparrow}} \nonumber \\
	&=& 2J_K^{+}J_K^{-} \int \frac{d^dK}{(2\pi)^d}\frac{n(\xi_{K,\uparrow}) - n(\xi_{K,\downarrow} ) }{2J_K^{z}} \nonumber \\
	&=& \frac{J_K^{+}J_K^{-}}{J_K^{z}}(n_{\uparrow} - n_{\downarrow})
\end{eqnarray}

Therefore, when the Kondo coupling is SU(2) symmetric  the mass terms cancel
and $\Pi^A + \Pi^B \approx J_K^2 \gamma_d |\omega|/q$ and thus
$\chi^{-1}(\vec{q},i\omega) = q^2 + \gamma_d J_K^2 \frac{|\omega|}{q}$,
where as usual we have neglected the linear in $\omega$ term because 
it is less relevant in the RG sense.  This special form of the bosonic
propagator has emerged in a number of other applications, the most
famous example being the gauge-fermion problem.  We will comment on
its consequence a little later.

With the inclusion of damping, the quadratic action now becomes:
\begin{eqnarray}
	\mathcal{S}_m 
		&=&	\int d\omega d^{d-1}q_{\perp} dq_{\parallel} \;
		m^{+}
		\left(q_{\perp}^2+b\frac{\omega}{q_{\perp} }\right)
		m^{-} \\
	\mathcal{S}_c 
		&=&	\int d\epsilon d^{d-1}k_{\perp} dk_{\parallel}\; 
		\bar{\psi}_{\sigma} 
		\left( i\epsilon - v_F k_{\parallel} -a_{\sigma} 
k_{\perp}^2 \right)
		\psi_{\sigma} 
\end{eqnarray}
where $a_{\sigma}$ and $b$ are simply couplings that control
the relative scaling between different
components of the action.  Their dimensions will be chosen 
to ensure the quadratic action is scale invariant. 
Significantly, in this $z_b=3$ theory the Berry phase no longer controls 
the dynamics, being instead overwhelmed by the damping term.
Physically, this is because the magnetization of the local moment
system is no longer conserved by itself once it can exchange
spin flips with the conduction electrons.

The scaling analysis now needs to be redone.
\begin{eqnarray}
	\left[ \epsilon \right] &=& 1 \nonumber \\
	\left[ k_{\parallel} \right] &=& 1 \nonumber\\
	\left[ k_{\perp} \right] &=& 1/z_b = 1/3 \nonumber\\
	\left[ a \right] &=& 1-1/z_b =2/3 \nonumber \\
	\left[ b \right] &=& 0 \nonumber \\
	\left[ \psi \right] &=& -(3z_b+d-1)/(2z_b) = -(8+d)/6 \nonumber\\
	\left[ m \right] &=& -(2z_b+d+1)/(2z_b) = -(7+d)/6
\end{eqnarray}
Note that in principle $a_{\sigma}$ and $\psi_{\sigma}$ could scale 
differently for different spin projections, but because of the way
they enter the action, we scale them identically.
With these choices, all the terms in the quadratic action are scale invariant.
The Kondo coupling terms,
\begin{eqnarray}
	\mathcal{S}_K^{\pm} &=& J_K^{\pm}
		\int d^{d-1} q_{\perp}dq_{\parallel} d\omega  d^{d-1} k_{\perp}dk_{\parallel} d\varepsilon \nonumber \\
&&\times		\Big[ 
			 \bar{\psi}_{k+q,\uparrow}\psi_{k,\downarrow}m^{-}_{q} +
			 \bar{\psi}_{k+q,\downarrow}\psi_{k,\uparrow}m^{+}_{q}
			\Big] 
			\\
	\mathcal{S}_K^{z} &=& J_K^z
		\int d^{d-1} q_{1\perp}dq_{1\parallel} d\omega_1
		d^{d-1} q_{2\perp}dq_{2\parallel} d\omega_2
		d^{d-1} k_{\perp}dk_{\parallel} d\varepsilon \; 
\nonumber\\
		&& \qquad \times \Big[
			 \bar{\psi}_{k+q_1-q_2,\uparrow}\psi_{k,\uparrow} m^{+}_{q_1}m^{-}_{q_2} 
+		 \bar{\psi}_{k+q_1-q_2,\downarrow}\psi_{k,\downarrow} m^{+}_{q_1}m^{-}_{q_2}
			\Big] \nonumber \\
\end{eqnarray}
are easily analyzed:
\begin{eqnarray}
	[ \mathcal{S}_K^{\pm} ] &=& 0  \nonumber \\
		&=& [J_K^{\pm}] +2[d^{d-1}k_{\perp}dk_{\parallel} d\varepsilon] + 2[\psi] + [m] \nonumber \\
		&=& [J_K^{\pm}] + 2\frac{d-1+2z_b}{z_b}-2\frac{3z_b+d-1}{2z_b} \nonumber \\
		&&-\frac{2z_b+d+1}{2z_b} 
		\nonumber \\
\implies [J_K^{\pm}] &=& (3-d)/(2z)
 \\
	\left[ \mathcal{S}_K^{z} \right] &=& 0  \nonumber \\
		&=& [J_K^z] +3[d^{d-1}k_{\perp}dk_{\parallel} d\varepsilon] + 2[\psi] + 2[m] \nonumber \\
		&=& [J_K^z] + 3\frac{d-1+2z_b}{z_b}-2\frac{3z_b+d-1}{2z_b} \nonumber \\
		&&-2\frac{2z_b+d+1}{2z_b} \nonumber \\
\implies [J_K^z] &=& (1-d)/z
\end{eqnarray}
The inclusions of $\omega/q$ damping into the quadratic part of the boson action has the effect of changing the dynamics from $z_b=2$ to $z_b=3$, however, there is no change to the dimension of the spin-flip Kondo coupling.  The longitudinal Kondo coupling is irrelevant for any $d>1$.

It turns out that a proper analysis of the fixed point 
requires insertion of the fermion self energy as well~\cite{Altshuler94_SI},
which we turn to next.

\section{Electron Self Energy and Non-Fermi Liquid Behavior}
In addition to the scaling analysis, we have another reason to determine
the electron self-energy. Anticipating that the non-Fermi liquid
contribution from the Kondo coupling to the magnons will be cut off
at the energy of order $\omega \sim \omega_c \sim  (I/W^2)\Delta^2$,
we wish to ascertain the magnitude
of the non-Fermi liquid term at this cutoff scale. This will allow
us to compare this term with some background Fermi liquid contributions.
Since the Kondo coupling also occurs in the modified magnon propagator,
we present here the calculation of the electron self-energy in some detail.

The leading order contribution to the electron 
self energy in $d=2$ is given 
by the dressed boson, bare fermion and no vertex correction, 
as depicted in figure~\ref{fig:self-energies}.
\begin{eqnarray}
	\Sigma_{\bar{\sigma}}(\vec{K},i\epsilon) 
	&=& J_K^{2} \int\frac{d^2q d\omega}{(2\pi)^3} 
		G^0_\sigma(\vec{K}+\vec{q},i\epsilon_m+i\omega_n)\chi(\vec{q},i\omega_n) 
\nonumber\\
	&=& J_K^{2}\int\frac{d^2q d\omega}{(2\pi)^3} 
		\frac{1}{i\epsilon+i\omega - \xi_{K+q,\sigma}}\;\frac{1}{q^2 - \Pi(\vec{q},i\omega) } 
\nonumber
\\
	&=& J_K^{2}\int\frac{d^2q d\omega}{(2\pi)^3} 
		\frac{1}{i\epsilon+i\omega - \xi_{K\sigma} - \frac{Kq}{m}\cos\theta} \nonumber\\
	&&\times \frac{1}{q^2 - \Pi(\vec{q},i\omega) }
\end{eqnarray}
From the previous section we have the result
$\Pi(\vec{q},i\omega_n) \approx -J_K^2\gamma\frac{|\omega|}{q}$.
For the integral over $\theta$ we use: 
$\int_0^{2\pi} \frac{1}{z+i \cos\theta} = \frac{2\pi\text{sgn}\Re (z)}{\sqrt{z^2+1}}$ for any complex $z$.
\begin{eqnarray}
	\Sigma_{\bar{\sigma}}(\vec{K},i\epsilon) 
	&=& J_K^{2}\int\frac{qdq d\omega}{(2\pi)^3} \frac{1}{Kq/m}\frac{1}{q^2 +J_K^2\gamma\frac{|\omega| }{q} } \nonumber\\
&&\times		\int d\theta\frac{1}{\frac{i\epsilon+i\omega - \xi_{K\sigma}}{\frac{Kq}{m}} - \cos\theta}
\nonumber \\
	&=& -iJ_K^{2}\int\frac{qdq d\omega}{(2\pi)^3} \frac{1}{Kq/m}\frac{1}{q^2 +J_K^2\gamma\frac{|\omega| }{q} } \nonumber\\
	&&\times
		\int d\theta\frac{1}{\frac{\epsilon+\omega + i \xi_{K\sigma}}{\frac{Kq}{m}} + i \cos\theta} 
\nonumber\\
	&=& -iJ_K^{2}\int\frac{qdq d\omega}{(2\pi)^3} \frac{1}{Kq/m}\frac{1}{q^2 +J_K^2\gamma\frac{|\omega| }{q} } \nonumber\\
	&&\times
		\frac{2\pi\,\text{sgn}( \epsilon + \omega)}{\sqrt{\left(\frac{\epsilon+\omega + i \xi_{K\sigma}}{\frac{Kq}{m}}\right)^2 + 1} } 
\end{eqnarray}
But in the regime of interest, and with the momentum restricted to $K \approx K_F$,  
we have $\frac{\epsilon+\omega + i\xi_{K\sigma}}{K_F/m} \ll 1$.
The self-energy then simplifies to
\begin{eqnarray}
	\Sigma_{\bar{\sigma}}(K_F,i\epsilon) 
	&\approx& -iJ_K^{2}\int\frac{qdq d\omega}{(2\pi)^3} \frac{1}{K_Fq/m}\frac{2\pi\,\text{sgn}( \epsilon + \omega)}{q^2 +J_K^2\gamma\frac{|\omega| }{q} } 
\nonumber\\
	&=& -J_K^{2}\frac{ im}{(2\pi)^2 K_F}\int_0^{\Lambda} dq \int_{-\infty}^{\infty}d\omega \frac{\text{sgn}( \epsilon + \omega)}{q^2 +J_K^2\gamma\frac{|\omega| }{q} } \nonumber \\
\end{eqnarray}
This integral is a little tricky.  First note that the frequency integral should have a cutoff, but this is complicated by the presence of the sgn function.  It would be incorrect to simply shift variables $\omega \to \omega+\epsilon$.  
The essential identity we need is:
\begin{eqnarray}
	\int_{-\Lambda}^{\Lambda} d\omega f(\omega) \text{sgn}(\omega+\epsilon)
	&=&
	2\int_0^{\epsilon} d\omega f(\omega)
\end{eqnarray}
which is only true for even functions: $f(\omega) = f(-\omega)$.  To see where this comes from, note first that for even functions:
\begin{eqnarray}
	\int_{a}^{b} d\omega f(\omega)
	&=&
	- \int_{-a}^{-b} d\omega f(\omega)
\nonumber
\end{eqnarray}
Next, to handle the sgn function we partition the integral into four regions:
\begin{eqnarray}
	\int_{-\Lambda}^{\Lambda} d\omega f(\omega) \text{sgn}(\omega+\epsilon)
	&=&
	-\int_{-\Lambda}^{-\epsilon} d\omega f(\omega) 
	+
	\int_{-\epsilon}^{0} d\omega f(\omega) \nonumber \\
	&&+
	\int_{0}^{\epsilon} d\omega f(\omega) +
	\int_{\epsilon}^{\Lambda} f(\omega) \nonumber
\end{eqnarray}
where the minus sign is the result of the sgn function.  Now we use the identity valid for even functions:
\begin{eqnarray}
	\int_{-\Lambda}^{\Lambda} d\omega f(\omega) \text{sgn}(\omega+\epsilon)
	&=&
	\int_{\Lambda}^{\epsilon} d\omega f(\omega) -
	\int_{\epsilon}^{0} d\omega f(\omega) \nonumber \\
	&& +
	\int_{0}^{\epsilon} d\omega f(\omega) +
	\int_{\epsilon}^{\Lambda} f(\omega) \nonumber \\
	&=&
	2\int_{0}^{\epsilon} d\omega f(\omega) \nonumber
\end{eqnarray}
Armed with this identity, the self energy is:
\begin{eqnarray}
	\Sigma_{\bar{\sigma}}(K_F ,i\epsilon) 
	&\approx& -J_K^{2}\frac{ 2im}{(2\pi)^2 K_F }\int_0^{\infty} dq \int_{0}^{\epsilon}d\omega \frac{1}{q^2 +J_K^2\gamma\frac{|\omega| }{ q } } \nonumber \\
	&=& -J_K^{2}\frac{ 2im}{(2\pi)^2 K_F J_K^2\gamma}\int_0^{\infty} dq q\log\left( 1 + J_K^2\frac{\gamma\epsilon}{q^3} \right)\nonumber \\
	&=& -\frac{ i2m}{(2\pi)^2 K_F \gamma} \frac{\pi}{\sqrt{3}}\left(J_K^2\gamma\epsilon \right)^{2/3}
\end{eqnarray}
Had we used a cutoff on the q-integral, we would have ended up with some unsightly hypergeometric functions whose
asymptotic form is the 
same as above, so it is easier to just set the cutoff to infinity straight
away.
For convenience, we have so far dropped the stiffness ($\rho_s$) factor
in the $q^2$ term of the boson propagator. Reintroducing this factor,
and taking $\rho_s \propto I$, we end up with the conduction electron self-energy quoted in the main text, 
Eq.~(7).

Redoing the calculations for $d=3$ is relatively straightforward, although now the integral will be UV divergent.  The only difference is that now we set $\vec{K}$ onto the x-axis since the $\phi$ variable is the one that runs from $0 \to 2\pi$.  This allows us to use the same identity on the $\phi$ integral that we used in the $d=2$ case for the $\theta$ integral.
\begin{eqnarray}
	\Sigma_{\bar{\sigma}}(\vec{K},i\epsilon) 
	&=& J_K^{2}\int\frac{q^2dq \sin\theta d\theta d\omega}{(2\pi)^4} \frac{1}{Kq/m}\frac{1}{q^2 +J_K^2\gamma\frac{|\omega| }{q} } \nonumber\\
	&&\times
		\int d\phi\frac{1}{\frac{i\epsilon+i\omega - \xi_{K\sigma}}{\frac{Kq}{m}} - \cos\phi} \nonumber \\
	&=& -iJ_K^{2}\int\frac{q^2dq \sin\theta d\theta d\omega}{(2\pi)^4} \frac{1}{Kq/m}\frac{1}{q^2 +J_K^2\gamma\frac{|\omega| }{q} } \nonumber\\
	&&\times
		\int d\phi\frac{1}{\frac{\epsilon+\omega + i \xi_{K\sigma}}{\frac{Kq}{m}} + i \cos\phi} \nonumber \\
	&=& -iJ_K^{2}\int\frac{q^2dq \sin\theta d\theta d\omega}{(2\pi)^4} \frac{1}{Kq/m}\frac{1}{q^2 +J_K^2\gamma\frac{|\omega| }{q} } \nonumber\\
	&&\times
		\frac{2\pi\,\text{sgn}( \epsilon + \omega)}{\sqrt{\left(\frac{\epsilon+\omega + i \xi_{K\sigma}}{\frac{Kq}{m}}\right)^2 + 1} } \nonumber
\end{eqnarray}
Within the regime of interest this simplifies to
\begin{eqnarray}
	\Sigma_{\bar{\sigma}}( K_F ,i\epsilon) 
	&\approx& -J_K^{2}\frac{2im}{(2\pi)^3K_F}\int qdq \int_0^{\epsilon} d\omega \frac{1}{q^2 +J_K^2\gamma\frac{|\omega| }{q} } \nonumber \\
	&=& \frac{-2J_K^{2}im}{(2\pi)^3K_F J_K^2\gamma}\int_0^{\Lambda} dq q^2 \log\left( 1 + \frac{J_K^2\gamma \epsilon}{q^3} \right) \nonumber \\
	&=& -\frac{2im}{(2\pi)^3K_F \gamma}\Bigg[
	\Lambda^3 \log\left(1+\frac{J_K^2\gamma\epsilon}{\Lambda^3} \right) \nonumber \\
	&&+ J_K^2\gamma\epsilon\log\left(1+\frac{\Lambda^3}{J_K^2\gamma\epsilon} \right)
	 \Bigg] \nonumber \\
	&\approx& -\frac{2im}{(2\pi)^3K_F \gamma}\Bigg[
		J_K^2\gamma - J_K^2\gamma\log\epsilon \nonumber \\
	&& + J_K^2\gamma\log\frac{\Lambda^3}{J_K^2\gamma}
	 \Bigg]\epsilon + O(\epsilon^2) 
\end{eqnarray}
So the leading singularity in $d=3$ is:
\begin{eqnarray}
	\Sigma
	&\propto& iJ_K^{2}\epsilon\log\epsilon
\end{eqnarray}
Again, recovering the stiffness factor leads to the form of the
conduction electron self-energy presented in the main text, 
Eq.~(7).

Holstein, Norton, and Pincus were the first to show that the transverse
electromagnetic field coupling remains unscreened and can in principle 
lead to non-Fermi liquid
behavior~\cite{Holstein73_SI}.
For a real electromagnetic field, the smallness
of the fine structure constant suppresses this effect
to extremely low temperatures.
Related non-Fermi liquid form appears in the gauge-fermion 
problem~\cite{Lee89_SI, Polchinski93_SI, Altshuler94_SI}.
More recently, similar self energies have been found near
quantum critical points and the nematic fermi 
fluid~\cite{Rech06_SI, Efremov08_SI, Oganesyan01_SI}.
The prevalence of this self energy results from the generic presence
of a massless $z_b=3$ boson coupled to a system with a Fermi surface.
The problem we have considered here 
has some important formal differences
from the gauge-fermion and critical Fermi liquid cases,
even in the $z_b=3$ continuum regime. 
One difference is in
the mechanism by which the boson propagators are gapless.
In the gauge-fermion problem,
gauge invariance guarantees the cancellation
of the mass term upon adding the bubble and tadpole diagrams 
in a large-N calculation
of the self energy of the vector potential~\cite{Tsvelik_book_SI}.
At the ferromagnetic QCP, the divergence of the correlation 
length ($\xi^{-2} \to 0$)
leads to gapless quantum critical fluctuations. In our case, 
it is the SU(2) spin symmetry
of the Kondo interaction which dictates that the
contribution from the longitudinal channel 
exactly cancels
that from the transverse channel.
A similar effect from the longitudinal mode of the ordered
itinerant antiferromagnet was
recently discussed by~\cite{Vekhter04_SI},
and we suspect that the cancellation
argument we advance here may apply to their case as well.
Another feature that is unique to our problem 
corresponds to the specific non-linear terms
[Eq.~\ref{non-linear}] that occur here, which
come into play in our RG analysis.
We have shown that these terms, while relevant 
for the pure Heisenberg problem, become irrelevant 
when the Kondo coupling to the fermions is
introduced.

We now turn to how the self-energy correction to 
fermions modify the damping term in the QNL$\sigma$M given 
in Eq.~(\ref{qnlsm-damping}). The damping remains to have 
the $\omega/q$ form. For the regime of our interest here,
$\omega \sim q^3$, both the self-energy and vertex corrections to 
the damping term are negligble. For generic $|\omega| \ll q$,
the self-energy and vertex corrections cancel with each other leaving
a subleading contribution~\cite{Altshuler94_SI, Kim94_SI}.

We close this section by noting that, even though we are deep in the
ferromagnetically ordered region, the phase space involved in the regime
we are considering is similar to that of fermions coupled to $z=3$ 
ferromagnetic quantum critical fluctuations. Parallel to the calculations
in the latter case, the temperature dependences of the electrical
resistivity and specific heat have the non-Fermi liquid form
given in the main text.

\section{Scaling with fully dressed propagators}
Now that we have the expression for the electron self energy
we can finally incorporate it into the fixed point and
redo the scaling analysis.
\begin{eqnarray}
	\mathcal{S}_m 
		&=&	\int d\omega d^{d-1}q_{\perp} dq_{\parallel} \;
		m^{+}
		\left(q_{\perp}^2+b\frac{\omega}{q_{\perp} }\right)
		m^{-} \\
	\mathcal{S}_c 
		&=&	\int d\epsilon d^{d-1}k_{\perp} dk_{\parallel}\; 
		\bar{\psi}_{\sigma} 
		\left( |\epsilon|^{d/z_b} - v_F k_{\parallel} -a_{\sigma} k_{\perp}^2 \right)
		\psi_{\sigma} \nonumber \\
\end{eqnarray}
Note that the self energy correction to the fermion in $d=3$ is actually $\epsilon\log\epsilon$, but for the purposes of scaling we can simultaneously treat the cases $d=2$ and $d=3$ by analyzing the form $\epsilon^{d/z_b}$.
To make every term in the quadratic action scale invariant we make the assignments:
\begin{eqnarray}
	\left[ k_{\perp} \right] &=& 1/d
\nonumber\\
	\left[ k_{\parallel} \right] &=& 1
\nonumber\\
	\left[ \epsilon \right] &=& z_b/d = 3/d
\nonumber\\
	\left[ a_{\sigma} \right] &=& 1-2/d
\nonumber\\
	\left[ \psi \right] &=& -(3d+z_b-1)/(2d) = -(3d+2)/(2d)
\nonumber\\
	\left[ m \right] &=& -(2d+z_b+1)/(2d) = -(2d+4)/(2d)
\end{eqnarray}
Inserting these dimensions into the Kondo coupling produces:
\begin{eqnarray}
	\left[ J_K^{\pm} \right] &=& (3-z_b)/(2d) = 0\\
	\left[ J_K^z \right] &=& (3-z_b-d)/d = -1
\end{eqnarray}
In both $d=2$ and $d=3$, we find that the insertion of the self energies has led to the marginality of the transverse Kondo coupling, and the irrelevance of the longitudinal channel.  This demonstrates that with the correct self energies built into the theory, which references the appropriate stable fixed point, there is never any unstable flow of the Kondo coupling.  The ferromagnetic phase with a small Fermi surface is stable to the Kondo coupling.

Parenthetically, note that the magnon scattering term
scales like:
\begin{eqnarray}
	\mathcal{S}_m^{(4)} &\sim& g\int \left(d^{d-1}q_{\perp}dq_{\parallel}d\omega \right)^3 (q_{\perp} m)^4 \\
\implies \left[g \right] &=& -\frac{3(d-1+d+z_b) + 4 - 2(2d+z_b+1)}{d}
\nonumber \\
				&=& \frac{1-z_b-2d}{d} 
\nonumber\\
				&=& -2\frac{d+1}{d}
\end{eqnarray}
which is always irrelevant.

\section{The effect of the cutoff}
Below the cutoff, $\omega< \omega_c \sim 
(I/W^2)\Delta^2$
and $q<q_c\sim (K_F/W)\Delta$,
the transverse Kondo coupling becomes irrelevant in the RG sense 
due to
phase space 
restrictions. The longitudinal Kondo coupling,
having the scaling dimension $(1-d)/z_b$, is irrelevant as well.
The non-Fermi liquid effect will therefore be cut off in this range.

To ascertain the strength of the non-Fermi liquid contribution,
we can compare the continuum contribution to the self energy,
Eq.~(7) of the main text,
with the background Fermi liquid contribution at the cutoff 
frequency $\omega_c$.
Adding a Coulomb interaction $u$ among the 
conduction electrons leads to a Fermi-liquid contribution to
the self-energy of the order 
$\Sigma_{FL}(\epsilon) \sim u^2\rho_0^3 \epsilon^2$. In $d=2$ 
we have
\begin{eqnarray}
	\Sigma_{NFL}(\epsilon \sim \omega_c) &\sim& 
(\rho_0J_K^4/I^2)^{1/3}\omega_c^{2/3} \sim J_K^{8/3}/W^{5/3} \nonumber\\ 
	&&\\
	\Sigma_{FL}(\epsilon\sim \omega_c) &=& u^2 \rho_0^3 \omega_c^2
\sim (u^2I^2/W^7)J_K^4
\end{eqnarray}
In the parameter range we consider, $J_K \ll |I| \ll W$,
$\Sigma_{NFL}(\epsilon \sim \omega_c)$ is much larger 
than $\Sigma_{FL}(\epsilon \sim \omega_c)$.
Note that in three dimensions, 
$\Sigma_{NFL}(\epsilon \sim \omega_c) \sim
\rho_0J_K^2 \omega_c / I \sim J_K^4/W^3$,
leading to a similar conclusion.

\section{Absence of Loop Corrections}
\subsection{Vertex Corrections}
For problems involving forward scattering of conduction electrons,
the inability of vertex corrections to qualitatively modify leading 
order results
has been established in related problems by a numbers of authors
~\cite{Polchinski93_SI, Altshuler94_SI, Kim94_SI, Yamamoto07_SI}.
The essence of the argument is a sort of Migdal's theorem 
reminiscent of the suppression of vertex
corrections in the electron-phonon problem~\cite{AGD_SI}.
Previous work utilized a large number of fermion flavors, but
we will take a slightly different approach which is more in line
with the spirit of the fermionic RG and,
like the original work by Migdal, focuses more explicitly
on kinematics and phase space.  The conclusions are
essentially the same.
The small parameter in our problem is 
$\Lambda/K_F \equiv 1/N_{\Lambda}$ 
which we use
to define the large-$N_{\Lambda}$ expansion.
(This $N_{\Lambda}\rightarrow \infty$ limit corresponds to asymptotically
low energies, {\it i.e.}, with the fermions approaching 
the Fermi surface.)
Denoting the number of loops by $L$, 
the structure of the beta function is given by:
\begin{eqnarray}
	\beta(J_K) 
	&=& b_0 J_K + \frac{d}{d \log s} \sum_{L=1}^{\infty} 
\frac{b_L(s)}{N_{\Lambda}^{e(L,d)} } J_K^{2L+1} \nonumber \\
	&\equiv& b_0 J_K + \frac{d}{d \log s} \Delta J_K
\end{eqnarray}
where loop integrals are performed over shells of width 
$\Lambda - \Lambda/s \approx \Lambda \log s$
with scaling parameter $s \equiv e^{\ell} \gtrapprox 1$.
$L$ is equal to the number of integrations needed to 
compute the diagram.
If the exponents $e(L,d)$ are positive for all
values of $L$ and $d$ ($>1$), the beta function is given by the
tree-level result ($b_0J_K$) in the large-$N_{\Lambda}$ limit, which
means vertex corrections can be neglected.
Since we have already shown that $b_0=0$, this would imply marginality 
to all orders.  The goal of this section is to demonstrate that this
is indeed the case.

\begin{figure*}
   \centering
   \includegraphics[width=2.5in]{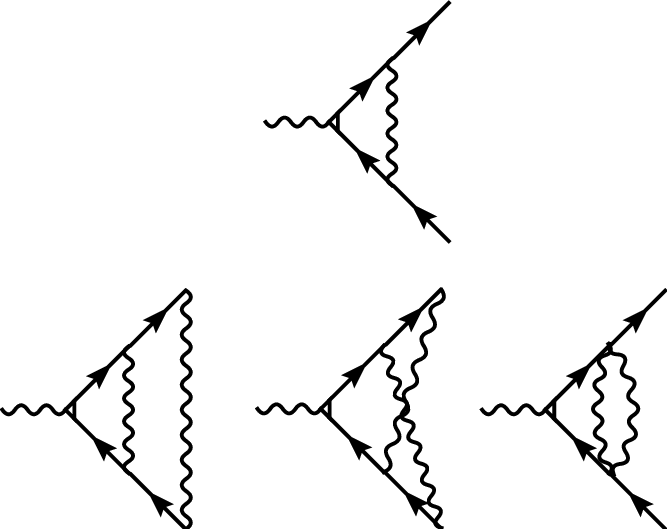}
   \caption{$L=1$ and $L=2$ vertex corrections. All propagators are dressed.}
   \label{fig:vertexCorr}
\end{figure*}

In what follows, we give two general arguments that demonstrate
that vertex corrections become increasingly suppressed in the loop expansion.
Specifically, a diagram with $L$-loops will come with a factor of $1/N_{\Lambda}^{L(d+3)/3}$,
\textit{i.e.} $e(L,d) = L(d+3)/3$.
We also illustrate the principle by calculating an example $L=1$ diagram
to demonstrate how this factor emerges.
We work with cutoffs in units of $K_F$ so that $N_{\Lambda} \equiv 1/\Lambda$.

The first argument is essentially just power-counting.
Every loop integral will introduce a factor of $\Lambda^{d+1}$ from
the measure of integration.  For $L$ loops, there will
be $2L$ fermion propagators (see Fig~\ref{fig:vertexCorr}) each carrying a factor
of $\Lambda^{-d/z}$ with $z=3$.  There will also be $L$
boson propagators which, because of the $\omega/q$ form
of the boson self energy, scale like $O(1)$.  
Thus, each diagram with $L$-loops contributes the
following amount of phase space.
\begin{eqnarray}
\label{eq:vCorr}
	\Delta J_K
	 &=&
	\sum_{L=1}^{\infty} b_L(s) J_K^{2L+1}(\Lambda^{d+1})^L (\Lambda^{-d/z})^{2L}(\Lambda/\Lambda)^{L} \nonumber \\
	 &=&
	\sum_{L=1}^{\infty} b_L(s) J_K^{2L+1}\Lambda^{[d(1-2/z)+1]L} 
\nonumber \\
	 &=&
	\sum_{L=1}^{\infty} \frac{b_L(s)}{N_{\Lambda}^{(d+3)L/3}} J_K^{2L+1}
\end{eqnarray}
Therefore $e(L,d) = (d+3)L/3 >0$, vertex corrects are kinematically suppressed,
and the tree level result (marginality) is the entire story.

\begin{figure*}
   \centering
   \includegraphics[width=2.5in]{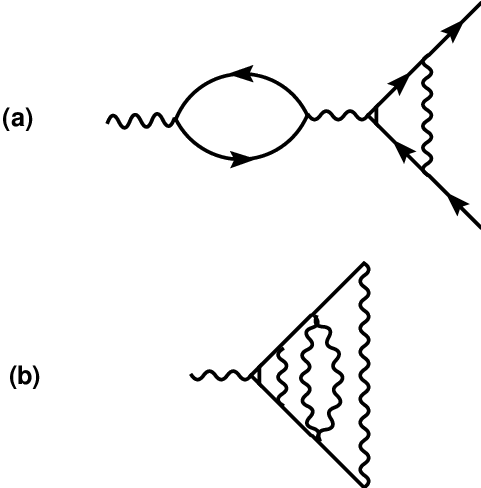}
   \caption{(a) Diagram which is not included because self energy corrections are already built
   into the dressed propagators.  (b) An example of $L=4$ diagram with 6 fermion propagators
   and 4 boson propagators.}
   \label{fig:moreVertexCorr}
\end{figure*}

The careful reader will have noticed that other classes of diagrams
are possible.  For example, Fig~\ref{fig:moreVertexCorr}a
shows a self-energy insertion into the boson propagator.
Iterates of diagrams like this might at first appear to compensate
for some powers of $N_{\Lambda}$ due to the pure fermion loops.
However, since we are using fully dressed propagators, this would
be double counting.  Such terms are already included by defining
the fixed point action to have the $\omega/q$ self energy from the beginning.

Another class of diagram is represented in Fig~\ref{fig:moreVertexCorr}b,
which is $L=4$ and with propagator powers of
$G^6\chi^4 = G^{2L-2}\chi^L$.  More generally, an exhaustive classification
of diagrams at order $L$, with $L$ even, will actually have $L/2+1$ subclasses which 
have factors
\begin{eqnarray}
	G^{2L}\chi^L, G^{2L-2}\chi^L, G^{2L-4}\chi^L,\cdots,G^L\chi^L
\nonumber 
\end{eqnarray}
If $L$ is odd, the series will terminate at $L+1$ rather than $L$,
and there will be $(L+1)/2$ subclasses.
However, since these subclasses only differ by {\it smaller}
powers of $G$ than the $G^{2L}$ we considered above, 
it is easy to see that they will be subleading
compared to the estimate given in equation~\ref{eq:vCorr}.

The second way to obtain Migdal's theorem more closely mirrors 
the antiferromagnetic case~\cite{Yamamoto07_SI}
and the ``leap to all loops'' of
the pure fermion problem~\cite{Shankar94_SI}.  
Begin by writing the quadratic parts of the action and rescaling
all momenta and energies by $\Lambda$ so the limits of integration
become dimensionless: $k \to \Lambda k$, $\epsilon \to \Lambda \epsilon$, etc.
\begin{eqnarray}
	\mathcal{S}_c 
	&=& \Lambda^{d+1}\sum^{N_{\Lambda}} \int d^dkd\epsilon \;
	\psi^{\dagger}(i\epsilon^{d/z}\Lambda^{d/z} - \Lambda k_{\parallel} - \Lambda^2 k_{\perp}^2 )\psi \nonumber\\
	&&\\
	\mathcal{S}_m
	&=&  \Lambda^{d+1}\int d^dq d\omega \;
	m^{+}(\Lambda^2 q^2 + \omega/q )m^{-}
\end{eqnarray}
For simplicity, we have omitted some prefactors.
To leading order in $1/N_{\Lambda}$ (small $\Lambda$), the dominant term in the fermionic part
is $\epsilon^{d/z}$ ($\epsilon\log\epsilon$ in $d=3$), 
while the $\omega/q$ term is largest in the bosonic part.  We therefore
rescale fields according to these terms, obtaining:
\begin{eqnarray}
	\psi \to \Lambda^{-(4d/3+1)/2} \psi \nonumber \\
	m^{\pm} \to \Lambda^{-(d+1)/2} m^{\pm}
\end{eqnarray}
This allows us to estimate the phase space contribution of the interaction term.
Rescaling according to this procedure, the Kondo coupling is given by:
\begin{eqnarray}
	&&J_K  \int \Lambda^{2(d+1)} d^dk d\epsilon d^dq d\omega
	\Lambda^{-(4d/3+1)} \psi^{\dagger} \psi \Lambda^{-(d+1)/2} m^{\pm} \nonumber \\
	&&\propto J_K \Lambda^{(d+3)/6}
\end{eqnarray}
Associated with every power of $J_K$ is a factor $1/N_{\Lambda}^{(d+3)/6}$,
and within the loop expansion the $L^{th}$-order correction
is given by $\Delta J_K/J_K \propto J_K^{2L} \propto 1/N_{\Lambda}^{(d+3)/3}$,
or $e(L,d) = (d+3)/3>0$, which is the same result we found earlier.
Therefore vertex correction can be neglected and the tree-level
result is asymptotically exact to all orders.

Note that the analog of this field rescaling for the pure fermion problem
results in a four-fermion coupling given by $\Lambda u \int \psi^4$.
In this case, Shankar found that the four-fermion coupling is still marginal 
despite the additional factor of $\Lambda$ induced by the field rescaling.
We are simply to regard $\Lambda$ as a small parameter (in units of $K_F$),
not a running variable.  Within the momentum shell approach, the beta function
is determined by finding the dependence on the parameter $s = e^{\ell}$
and computing the derivative $d/d\ell$, not by finding any explicit
dependence on $\Lambda$ as is done in the field theory approach.

We have now proven that vertex corrections can be neglected in the
large-$N_{\Lambda}$ limit.  To demonstrate how the peculiar
exponent $e(L,d)=(d+3)/3$ arises in a concrete example, let us calculate
the first, $L=1$, vertex correction shown in figure~\ref{fig:vertexCorr}.
\begin{eqnarray}
	\Delta J_K(\vec{p},i\epsilon; \vec{Q},i\Omega)
	&=& J_K^3 \int d^dq d\omega
	G(\vec{p}+\vec{q},i\epsilon+i\omega)\nonumber \\
	&&\times
	G(\vec{p}+\vec{q}+\vec{Q},i\epsilon+i\omega+i\Omega)
	\chi(\vec{q}, i\omega) \nonumber \\
\end{eqnarray}
We can set the fermionic variables $\vec{p}=0$ (measured from the 
patch origin) 
and $\epsilon = 0$ since any deviation would be irrelevant in the RG sense. 
In contrast, the variables $\vec{Q}$ and $\Omega$ belong to the external boson
which we keep nonzero, keeping in mind that our problem has cutoffs
$\omega_c$ and $q_c$.
\begin{widetext}
\begin{eqnarray}
	\Delta J_K(0,0; \vec{Q},i\Omega)
	&=& J_K^3
	\int d^dq d\omega
	\frac{1}{
	i\epsilon_0^{1/3}|\omega|^{2/3}
	- v_{\uparrow}q_{\parallel} 
	- v_{\uparrow}q_{\perp}^2/(2K_{F\uparrow})} 
	\frac{\sqrt{q_{\perp}^2 + q_{\parallel}^2} }{
	(q_{\perp}^2 + q_{\parallel}^2)^{3/2} 
	+ \gamma|\omega|} 
\nonumber \\
&&\times
	\frac{1}{
	i\epsilon_0^{1/3}|\omega+\Omega|^{2/3}
	- v_{\downarrow}(q_{\parallel} +Q_{\parallel} )
	-v_{\downarrow}(q_{\perp}^2+Q_{\perp}^2)/(2K_{F\downarrow})} 
\nonumber
\end{eqnarray}
\end{widetext}
One way to demonstrate Migdal's theorem is to factorize integrands
of momentum integrals according to a certain procedure,
as detailed by several authors
~\cite{Altshuler94_SI, Abanov03_SI, Chubukov05_SI}.
Physically, this relies of the fact that fermions are much faster than
bosons.  Formally, this can be accomplished by rescaling
$v_F \to N v_F$ and similarly for the coupling;
see Appendix A of of ref \cite{Chubukov05_SI}.

The validity of the factorization is not entirely obvious.
Within a large-$N$ treatment, a thorough analysis has been done where
numerical comparisons show that the factorization approximation only begins
to break down at relatively high temperatures~\cite{Abanov03_SI},
outside the regime we consider here.  In the next section, we 
show that the factorization of momentum integrations
applies in the large-$N_{\Lambda}$ limit (without invoking 
large-$N$).  For the rest of this section,
we first proceed with such a factorization.

In such a case,
the only parts of the integrand that depend on $q_{\perp}$ are the
bosonic propagators.  This allows us to define a momentum independent boson
given by the fully momentum dependent
propagator integrated along the Fermi surface:
\begin{eqnarray}
	\chi_1(i\omega; \Lambda_{\perp}, q_c) 
	&\equiv& \int d^{d-1}q_{\perp} 
	\chi(\vec{q},i\omega)|_{q_{\parallel}=0} \nonumber \\
	&=& \int d^{d-1}q_{\perp} 
	\frac{
	\Theta(q_{\perp} - q_{c\perp})
	\Theta(\Lambda_{\perp} - q_{\perp} )
	}{q_{\perp}^2 + \gamma\frac{|\omega|}{|q_{\perp} |} }\nonumber \\
\end{eqnarray}
Note, once again, that we adopt the convention of ref. \cite{Altshuler94_SI} in labeling
parallel and perpendicular components.  
Also note that unlike other problems, we have
a natural infrared regularization provided by the cutoff on
bosonic modes.  All integrals thus have both UV regularizations $\Lambda$
and IR regularizations $q_c$ and $\omega_c$.
Moreover, loop integrals will be performed over momentum and energy shells,
rather than extending the limits of integration to infinite intervals.
This is the reason why we do \textit{not} find
a non-analytic $q^{3/2}$ correction to the static
boson propagator, in contrast to theories
for the itinerant ferromagnetic quantum critical 
point~\cite{Chubukov04_SI}.

To leading order in large-$N_{\Lambda}$, the vertex correction can 
now be written in factorized form:
\begin{eqnarray}
	\Delta J_K(0,0; Q_{\parallel},i\Omega)
	&=& J_K^3
	\int d\omega \chi_1(i\omega; \Lambda_{\perp}, q_c) \nonumber \\
&&\times
	\int dq_{\parallel} 
	\frac{1}{
	i\epsilon_0^{1/3}|\omega|^{2/3}
	- v_{\uparrow}q_{\parallel} } \nonumber\\
&&\times	\frac{1}{
	i\epsilon_0^{1/3}|\omega+\Omega|^{2/3}
	- v_{\downarrow}(q_{\parallel}+Q_{\parallel}) } \nonumber
\end{eqnarray}
Note that the dimensional dependence is confined to $\chi_1(i\omega)$,
while the $q_{\parallel}$ dependence is isolated in the fermionic propagators.
The dependence on external
$Q_{\perp}$ has dropped out, which is higher order in $1/N_{\Lambda}$.
To proceed, we consider $d=2$ for this illustrative example.

The range of integration requires some comment.
Within the momentum-shell scheme,
each loop integral consists of a number of 
``slabs'' in phase space of width
$\Lambda - \Lambda/s^{\eta} \approx \eta \Lambda\log s$,
where $\eta$ is the scaling dimension of the appropriate direction.
Within each slab, the integrand can be approximated by its value
at the cutoff.
For example, at one-loop we can write
\begin{eqnarray}
	I
	&\equiv& \int d\omega dq_{\parallel} dq_{\perp} f(\omega, q_{\parallel}, q_{\perp}) \nonumber \\
	&\approx& 
	\left[  q_{\perp} \right] \Lambda_{\perp}\log s 
	\int d\omega dq_{\parallel} [ f(\omega, q_{\parallel}, \Lambda_{\perp}) + f(\omega, q_{\parallel}, -\Lambda_{\perp}) ] \nonumber \\
&&+ 
	\left[  q_{\parallel} \right] \Lambda_{\parallel}\log s 
	\int d\omega dq_{\perp} [ f(\omega, \Lambda_{\parallel}, q_{\perp}) + f(\omega, -\Lambda_{\parallel}, q_{\perp}) ] \nonumber \\
&&+ 
	\left[  \omega \right] \Lambda_{\omega}\log s 
	\int dq_{\parallel} dq_{\perp} [ f(\Lambda_{\omega}, q_{\parallel}, q_{\perp}) + f(-\Lambda_{\omega}, q_{\parallel}, q_{\perp}) ] \nonumber \\
	&\equiv& (\Lambda_{\perp}  I_{\perp}+ \Lambda_{\parallel} I_{\parallel}+ \Lambda_{\omega} I_{\omega})\log s
\end{eqnarray}
We have divided the loop integral into a sum of $(d+1)$ terms which represent the slabs
directed along each of the $(d+1)$ hyperplanes.
This is simply the multidimensional generalization of the trivial result:
$\int_{\Lambda/s^{\eta}}^{\Lambda}dx\; f(x) + \int^{-\Lambda/s^{\eta}}_{-\Lambda}dx\; f(x) \approx
\eta\Lambda\log s [f(\Lambda)+ f(-\Lambda)]$.
Let us consider one of these slab integrals.
\begin{eqnarray}
	\Lambda_{\omega} I_{\omega}
	&=& \left[ \omega \right] \Lambda_{\omega}J_K^3
	\int dq_{\perp} 
	\frac{1}{q_{\perp}^2 + \gamma\frac{\Lambda_{\omega} }{|q_{\perp}|} }
	\int dq_{\parallel}\nonumber\\
&&\times	\Big[
	\frac{1}{
		i\epsilon_0^{1/3}\Lambda_{\omega}^{2/3}
		- v_{\uparrow}q_{\parallel} } 
	\frac{1}{
		i\epsilon_0^{1/3}(\Lambda_{\omega}+\omega_c)^{2/3}
		- v_{\downarrow}(q_{\parallel}+q_{c}) } \nonumber \\
&&+		\frac{1}{
		i\epsilon_0^{1/3}\Lambda_{\omega}^{2/3}
		+ v_{\uparrow}q_{\parallel} } 
	\frac{1}{
		i\epsilon_0^{1/3}(\Lambda_{\omega}-\omega_c)^{2/3}
		+ v_{\downarrow}(q_{\parallel}+q_{c}) }
	\Big] \nonumber\\
\end{eqnarray}
where we have take the external frequency and moment down to the
cutoffs $q_c$ and $\omega_c$, 
and assumed $0<\omega_c < \Lambda_{\omega}$.
This integral is factorized, with the first factor being given by
\begin{widetext}
\begin{eqnarray}
	\chi_{1} (\omega = \Lambda_{\omega}) 
	&=& 
	\chi_{1} (\omega = -\Lambda_{\omega}) \nonumber \\
	&=&
	\int dq_{\perp} 
	\frac{1}{q_{\perp}^2 + \gamma\frac{\Lambda_{\omega} }{|q_{\perp}|} } 
	\nonumber \\
	&=& (1/3)(\gamma\Lambda_{\omega})^{-1/3}
	\Bigg[ 
		2\sqrt{3}\arctan\left( \frac{1-2q_c(\gamma\Lambda_{\omega})^{-1/3} }{ \sqrt{3} } \right) 
	-	2\sqrt{3}\arctan\left( \frac{1-2q_c(\gamma\Lambda_{\omega})^{-1/3} }{ \sqrt{3} } \right) \nonumber\\
&&	+	2\log(q_c+(\gamma\Lambda_{\omega})^{1/3})
	-	2\log(\Lambda_{\perp} + (\gamma\Lambda_{\omega})^{1/3}) 
	-	\log(q_c^2 - q_c (\gamma\Lambda_{\omega})^{1/3} + (\gamma\Lambda_{\omega})^{2/3}) \nonumber \\
&&	+	\log(\Lambda_{\perp}^2 - \Lambda_{\perp} (\gamma\Lambda_{\omega})^{1/3} + (\gamma\Lambda_{\omega})^{2/3})
	\Bigg] 
\end{eqnarray}
For an estimate of this factor, we must first take the limit $q_c \to 0$, since this
must be smaller than the UV cutoffs:
\begin{eqnarray}
	\lim_{q_c \to 0} \chi_{1} (\omega = \Lambda_{\omega}) 
	&=& (1/9)(\gamma\Lambda_{\omega})^{-1/3}
	\Bigg[ 
		\pi\sqrt{3} 
	-	6\sqrt{3}\arctan\left( \frac{1-2\Lambda_{\perp}(\gamma\Lambda_{\omega})^{-1/3} }{ \sqrt{3} } \right)
	-	6\log(\Lambda_{\perp} + (\gamma\Lambda_{\omega})^{1/3}) \nonumber\\
&&	+	3\log(\Lambda_{\perp}^2 - \Lambda_{\perp} (\gamma\Lambda_{\omega})^{1/3} + (\gamma\Lambda_{\omega})^{2/3})
	\Bigg]
\end{eqnarray}
Next we set $\Lambda_{\omega} = \Lambda_{\perp} = \Lambda$, 
then take a small $\Lambda$ expansion, finding:
\begin{equation}
	\chi_{1} 
	\propto
	\Lambda/\gamma - \frac{2}{5\gamma^2}\Lambda^3 + O(\Lambda^5)
\end{equation}.
The second factor is a more complicated integral.
\begin{eqnarray}
	\int (GG|_{\Lambda_{\omega}} +GG|_{-\Lambda_{\omega}})  &=&
	\int dq_{\parallel}
	\Big[
	\frac{1}{
		i\epsilon_0^{1/3}\Lambda_{\omega}^{2/3}
		- v_{\uparrow}q_{\parallel} } 
	\frac{1}{
		i\epsilon_0^{1/3}(\Lambda_{\omega}+\omega_c)^{2/3}
		- v_{\downarrow}(q_{\parallel}+q_{c}) } \nonumber \\
&&+		\frac{1}{
		i\epsilon_0^{1/3}\Lambda_{\omega}^{2/3}
		+ v_{\uparrow}q_{\parallel} } 
	\frac{1}{
		i\epsilon_0^{1/3}(\Lambda_{\omega}-\omega_c)^{2/3}
		+ v_{\downarrow}(q_{\parallel}+q_{c}) }
	\Big] \nonumber \\
&=&
		\frac{1}
		{i\epsilon_0^{1/3}[(\Lambda_{\omega}+\omega_c)^{2/3}
		-\Lambda_{\omega}^{2/3}] -v_F q_c
		} \nonumber \\
&&\times
	\Bigg\{
		-\frac{2i}{v_{\uparrow}}
		\Big[
			\arctan
			\left( 
				\frac{v_{\uparrow}\Lambda_{\parallel} }{\epsilon_0^{1/3}|\Lambda_{\omega}|^{2/3}} 
			\right)
		-	\arctan
			\left( 
				\frac{v_{\uparrow}q_{c} }{\epsilon_0^{1/3}|\Lambda_{\omega}|^{2/3}} 
			\right)
		\Big] \nonumber \\
&&		-\frac{1}{v_{\downarrow} } \Big(
		\frac{1}{2} \log
		\frac
		{\epsilon_0^{2/3}|\omega_c+\Lambda_{\omega}|^{4/3}+v_{\downarrow}^2(q_{c}-\Lambda_{\parallel})^2}
		{\epsilon_0^{2/3}|\omega_c+\Lambda_{\omega}|^{4/3}+v_{\downarrow}^2(q_{c}+\Lambda_{\parallel})^2} 
-		\frac{1}{2} \log
		\frac
		{\epsilon_0^{2/3}|\omega_c+\Lambda_{\omega}|^{4/3}}
		{\epsilon_0^{2/3}|\omega_c+\Lambda_{\omega}|^{4/3}+4v_{\downarrow}^2 q_{c}^2} \nonumber \\
&&		+i\arg [i \epsilon_0^{1/3}|\omega_c+\Lambda_{\omega}|^{2/3} - v_{\downarrow}(q_{c}-\Lambda_{\parallel})] 
		-i\arg [i \epsilon_0^{1/3}|\omega_c+\Lambda_{\omega}|^{2/3} - v_{\downarrow}(q_{c}+\Lambda_{\parallel})] \nonumber \\
&&		+i\arg [i \epsilon_0^{1/3}|\omega_c+\Lambda_{\omega}|^{2/3}] 
		-i\arg [i \epsilon_0^{1/3}|\omega_c+\Lambda_{\omega}|^{2/3} - 2v_{\downarrow}q_{c}]
		\Big)
	\Bigg\} \nonumber \\
&&+
		\frac{1}
		{i\epsilon_0^{1/3}[(\Lambda_{\omega}-\omega_c)^{2/3}
		-\Lambda_{\omega}^{2/3} - v_F q_c]
		} \nonumber \\
&&\times
	\Bigg\{
		-\frac{2i}{-v_{\uparrow}}
		\Big[
			\arctan
			\left( 
				\frac{-v_{\uparrow}\Lambda_{\parallel} }{\epsilon_0^{1/3}|\Lambda_{\omega}|^{2/3}} 
			\right)
		-	\arctan
			\left( 
				\frac{-v_{\uparrow}q_{c} }{\epsilon_0^{1/3}|\Lambda_{\omega}|^{2/3}} 
			\right)
		\Big] \nonumber \\
&&+		\frac{1}{v_{\downarrow} } \Big(
		\frac{1}{2} \log
		\frac
		{\epsilon_0^{2/3}|-\omega_c+\Lambda_{\omega}|^{4/3} -v_{\downarrow}^2(q_{c}-\Lambda_{\parallel})^2}
		{\epsilon_0^{2/3}|-\omega_c+\Lambda_{\omega}|^{4/3} -v_{\downarrow}^2(q_{c}+\Lambda_{\parallel})^2} \nonumber \\
&&-		\frac{1}{2} \log
		\frac
		{\epsilon_0^{2/3}|-\omega_c+\Lambda_{\omega}|^{4/3}}
		{\epsilon_0^{2/3}|-\omega_c+\Lambda_{\omega}|^{4/3}-4v_{\downarrow}^2 q_{c}^2} \nonumber \\
&&		+i\arg [i \epsilon_0^{1/3}|-\omega_c+\Lambda_{\omega}|^{2/3} +v_{\downarrow}(q_{c}-\Lambda_{\parallel})] \nonumber \\
&&		-i\arg [i \epsilon_0^{1/3}|-\omega_c+\Lambda_{\omega}|^{2/3} +v_{\downarrow}(q_{c}+\Lambda_{\parallel})] \nonumber \\
&&		+i\arg [i \epsilon_0^{1/3}|-\omega_c+\Lambda_{\omega}|^{2/3}] 
		-i\arg [i \epsilon_0^{1/3}|-\omega_c+\Lambda_{\omega}|^{2/3} + 2v_{\downarrow}q_{c}]
		\Big)
	\Bigg\}
\end{eqnarray}
\end{widetext}
where we have neglected terms of order $ v_{\uparrow} - v_{\downarrow} \sim \Delta/\mu$,
and used 
$\int dq_{\parallel} = \int dq_{\parallel} \Theta(q_{\parallel} - q_{c})\Theta(\Lambda_{\parallel}-q_{\parallel})$.
Using the same procedure as for the previous factor, as
well as the following simplifications,
$q_c/\Lambda \ll 1$, 
$\omega_c/\Lambda \ll 1$,
$\omega_c/q_c \ll 1$, 
and
$\omega_c/\omega \ll 1$, 
we find the rather simple result:
\begin{eqnarray}
	\int (GG|_{\Lambda_{\omega}} +GG|_{-\Lambda_{\omega}})  
	&\propto& \Lambda^{1/3}\Lambda^{-2/3} = \Lambda^{-1/3}
\end{eqnarray}

Putting it all together, we find
\begin{eqnarray}
	\Lambda_{\omega} I_{\omega} 
	&\sim& \left[ \omega \right] \Lambda (\log s) \Lambda \Lambda^{-1/3} 
	\nonumber \\
	&=& \frac{3}{2} \Lambda^{5/3} (\log s)
\end{eqnarray}
By a similar analysis, the other slab contributions
can be shown to have the same exponent:
$\Lambda_{\perp} I_{\perp} \sim \Lambda^{5/3}$ and 
$\Lambda_{\parallel} I_{\parallel} \sim \Lambda^{5/3}$.
Therefore, the one-loop correction to the beta function is given by:
\begin{eqnarray}
	\delta J_K &\sim& \frac{d}{d\log s} \Lambda^{5/3} \log s 
	\nonumber \\
	&=& \Lambda^{5/3} 
	\nonumber \\
	&=& 1/N_{\Lambda}^{5/3}
\end{eqnarray}
which confirms our previous and more general derivations of Migdal's theorem:
$e(L=1,d=2) = (2+3)/3 = 5/3$.

To summarize, we have demonstrated Migdal's theorem in three different ways, including
an explicit calculation of the one-loop integral as a concrete example.

An interesting future direction would be to consider calculations of this sort with finite $\Lambda/K_F$,
akin to $1/N_{\Lambda}$ corrections.
In particular, it is easy to imagine that special bandstructures might possess 
Fermi surface features, such as nesting or van-Hove singularities, 
that might lead to significantly different conclusions.
For such cases, however, it would then be necessary to consider specific materials
with realistic bandstructures, and we would lose  our ability to make universal statements.  
For this reason we remain content with the $N_{\Lambda}=\infty$
limit which should be valid under generic circumstances, and leave to future work
detailed investigations of material-specific bandstructures where $1/N_{\Lambda}$
corrections might play an important role.
We also point out that identifying the $N_{\Lambda} = \infty$ theory is in itself
a non-trivial result.  After all, Landau Fermi liquid theory is the $N_{\Lambda}=\infty$
limit of the interacting fermion problem~\cite{Shankar94_SI} which has been profoundly useful
despite the fact that, by itself, $1/N_{\Lambda}$ corrections are not captured.

\subsection{Factorization of Momentum Integrals}
The property of $q_{\parallel}-q_{\perp}$ integrations 
has previously been discussed within a large-$N$ limit,
where $N$ is the number of fermion flavors~\cite{Abanov03_SI}.
These theories typically perform loop integrations over all 
of phase space, in which case
it becomes necessary to introduce
the large factor $N$ in order to properly weight the desired kinematic range.  Working
with cutoffs explicitly, as we do, the integrals are more difficult to compute without the
technology of residue calculus, however, the physical kinematic regime is more
naturally apparent.  Here, we demonstrate the validity of the factorization
approximation used in the previous section, but we do not require a large number
of fermion flavors.  Instead, our large parameter is the ratio $N_{\Lambda} \equiv K_F/\Lambda$.

\begin{figure*}
   \centering
   \includegraphics[width=6in]{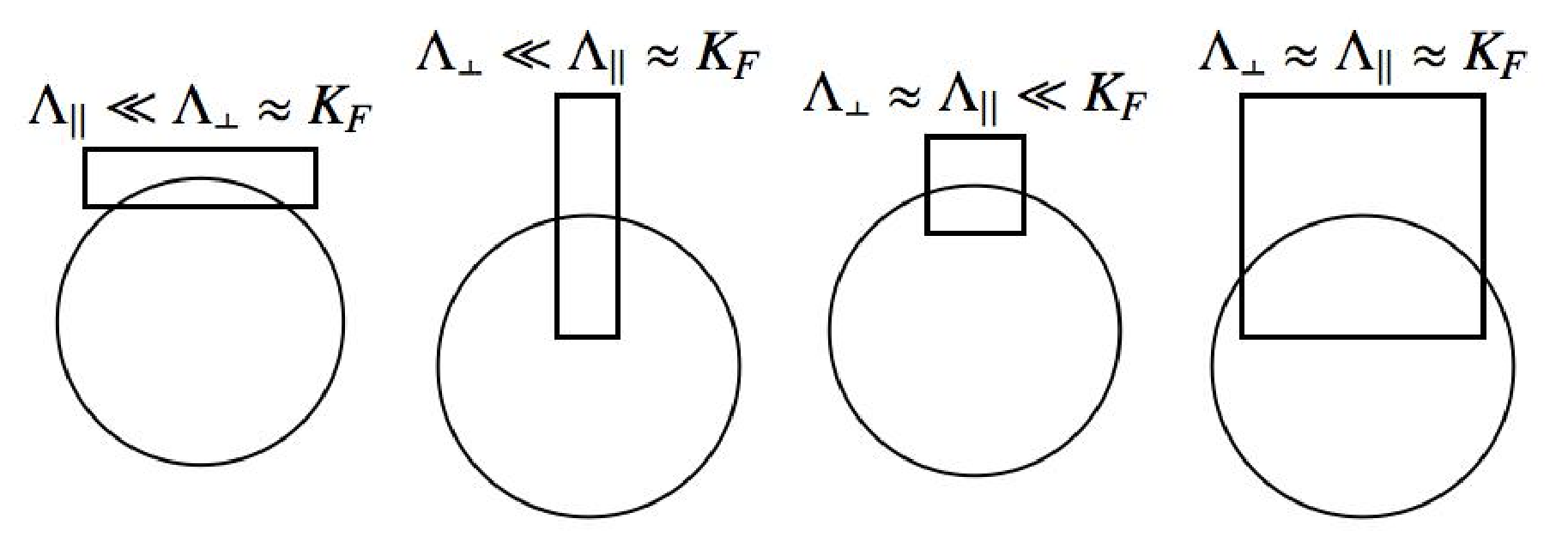}
   \caption{This figure depicts the various choices we have in choosing the size of
   our integration cutoffs in relation to each other, 
   and the scale set by the Fermi momentum $K_F$.
   To restrict to low-energy excitations we must have 
   $\Lambda_{\parallel} \ll K_F$.  To ensure that we have a large number
   of patches, we must insist that $\Lambda_{\perp} \ll K_F$.
   The most convenient and sensible choice is to take
   $\Lambda_{\perp} \sim \Lambda_{\parallel} \ll K_F$.  
   Even though the cutoffs are of similar size, we still have
   $q_{\perp} \gg q_{\parallel}$, as indicated by the next figure.}
   \label{fig:cutoffs}
\end{figure*}

Consider a low-energy fermion represented by a point infinitesimally near the Fermi surface.
This point on the Fermi surface defines the origin of our coordinate system.
Since this patch of surface is defined by its normal, we decompose the coordinate
system into components parallel and perpendicular to this normal vector.
A low energy, forward scattering excitation involving this state will be contained within a box
of size $\Lambda$ near this point of the Fermi surface, and we demand $\Lambda \ll K_F$.  
The momentum transfer between these two fermion states we label with 
$\vec{q} = (q_{\parallel},\vec{q}_{\perp})$.
The factorization approximation is valid in the 
limit where the dimension of the box along the Fermi surface is
much smaller than the Fermi wavevector: $\Lambda_{\perp} \ll K_F$.
There are three ways to choose a small cutoff, as depicted in Fig~\ref{fig:cutoffs}.
Either set 
$\Lambda_{\parallel} \ll \Lambda_{\perp} \sim K_F$,
or
$\Lambda_{\perp} \ll \Lambda_{\parallel} \sim K_F$,
or
$\Lambda_{\perp} \sim \Lambda_{\parallel} \ll K_F$.
The first choice might lead one to believe that
the number of patches is not large, which is not the case.
The second option appears to suggest that 
$q_{\perp} \ll q_{\parallel}$, which is opposite to the regime
we wish to consider.  Furthermore, it includes high-energy
excitations far from the mass shell.  
The third choice seems most natural, and it turns out to 
be the most convenient in terms of calculations as well,
as indicated in the previous section.
It might lead one to believe that the scaling is isotropic,
but we will show below that this is not the case.
Finally, a fourth possibility,
where
$\Lambda_{\perp} \sim \Lambda_{\parallel} \sim K_F$
has been used in calculations by other authors.
When calculating with the fourth option, where loop
integrals essentially extend to infinity,
it is necessary to 
rescale the Fermi velocity by a large factor such
as an artificially large number of fermion
flavors~\cite{Chubukov05_SI}.
The hope is that the $N=\infty$ results will
be connected to the $N=2$ case we wish to understand,
rather than the $N=0$ limit which is qualitatively 
different~\cite{Sedrakyan09_SI, Altshuler94_SI}.
We choose, instead, to rely on the fact that $ \Lambda \ll K_F$
which does not require us to resort to large-$N$, but only
large-$N_{\Lambda}$ which is simply the limit of the
low-energy field theory.

To see the small error made by the factorization approximation when
$\Lambda_{\perp} \sim \Lambda_{\parallel} \ll K_F$, 
consider the one-loop vertex correction we calculated in the previous section, with and
without the factorization approximation:
\begin{widetext}
\begin{eqnarray}
	\Delta J_K
	&=& J_K^3
	\int d^dq d\omega
	\frac{1}{
	i\epsilon_0^{1/3}|\omega|^{2/3}
	- v_{\uparrow}q_{\parallel} 
	- v_{\uparrow}q_{\perp}^2/(2K_{F\uparrow})} 
	\frac{\sqrt{q_{\perp}^2 + q_{\parallel}^2} }{
	(q_{\perp}^2 + q_{\parallel}^2)^{3/2} 
	+ \gamma|\omega|} 
	\nonumber \\
&&\times
	\frac{1}{
	i\epsilon_0^{1/3}|\omega+\Omega|^{2/3} 
	- v_{\downarrow}(q_{\parallel} +Q_{\parallel} )
	-v_{\downarrow}(q_{\perp}^2+Q_{\perp}^2)/(2K_{F\downarrow})} 
\\
	\Delta J_K\Big|_{\text{factorized}}
	&=& J_K^3
	\int d^dq d\omega
	\frac{1}{
	i\epsilon_0^{1/3}|\omega|^{2/3}
	- v_{\uparrow}q_{\parallel}} 
	\frac{ |q_{\perp}|  }{
	q_{\perp}^3
	+ \gamma|\omega|} 
	\frac{1}{
	i\epsilon_0^{1/3}|\omega+\Omega|^{2/3}
	- v_{\downarrow}(q_{\parallel} +Q_{\parallel})} 
\end{eqnarray}
\end{widetext}

The integrands are sharply peaked in phase space along surfaces
defined by the zeros of the inverse propagators.  For $\Delta J_K$
this corresponds to the surface defined by:
\begin{eqnarray}
&&	G^{-1}(q_{\perp}, q_{\parallel},i\omega)
	G^{-1}(q_{\perp}+Q_{\perp}, q_{\parallel}+Q_{\parallel},i\omega+i\Omega)\nonumber\\
&&\times
	\chi^{-1}(q_{\perp}, q_{\parallel}, i\omega)=0
\end{eqnarray}
while for $\Delta J_K\big|_{\text{factorized}}$, the surface is defined by:
\begin{eqnarray}
&&	G^{-1}(0, q_{\parallel},i\omega)
	G^{-1}(0, q_{\parallel}+Q_{\parallel},i\omega+i\Omega)\nonumber\\
&&\times
	\chi^{-1}(q_{\perp}, 0, i\omega)=0
\end{eqnarray}

\begin{figure*}
   \centering
   \includegraphics[width=6in]{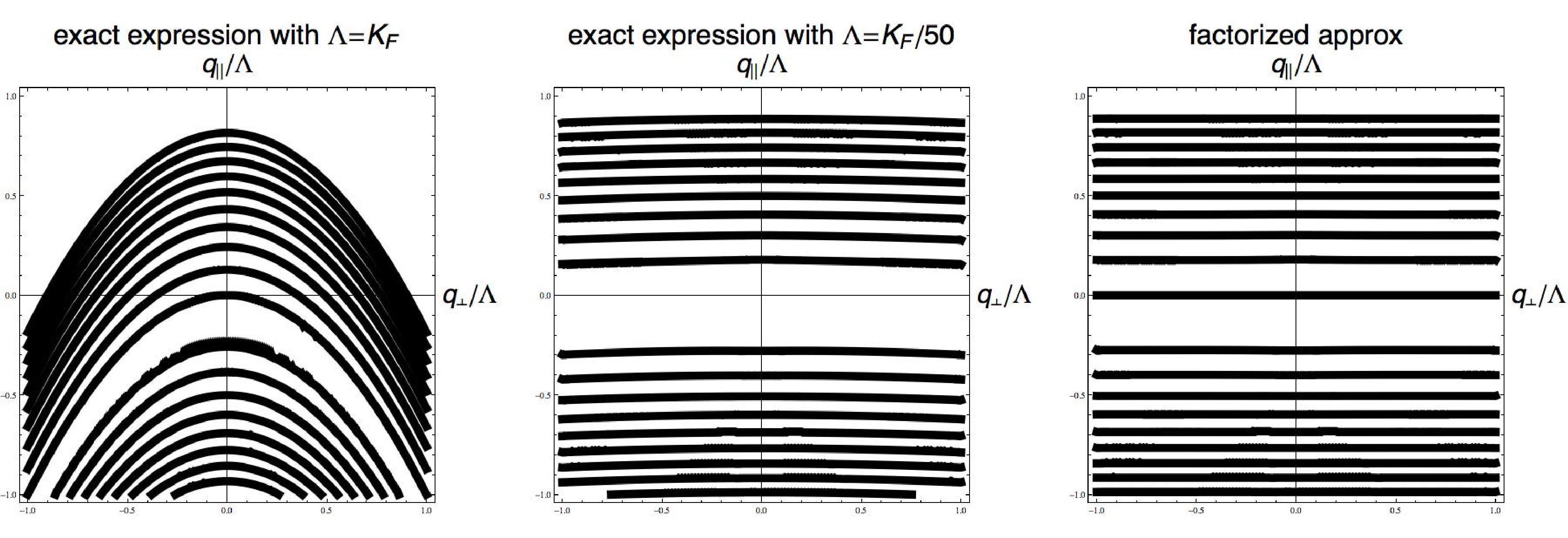}
   \caption{
   The left panel shows constant energy contours defined by the equation
   $	G^{-1}(q_{\perp}, q_{\parallel},i\omega)
	G^{-1}(q_{\perp}+Q_{\perp}, q_{\parallel}+Q_{\parallel},i\omega+i\Omega)
	\chi^{-1}(q_{\perp}, q_{\parallel}, i\omega)=0$,
   corresponding to the peaked regions of the
    unfactorized (``exact'') integrand of the 
   one-loop vertex correction with 
   the unphysical value $\Lambda = K_F$.
   The middle panel is the same exact expression, but
   with the more reasonable $\Lambda = K_F/50$.  
   Finally, the right panel
   depicts the constant energy contours of the highly
   peaked regions of the integrand using the 
   factorization approximation; these curves are
   defined by:
   $	G^{-1}(0, q_{\parallel},i\omega)
	G^{-1}(0, q_{\parallel}+Q_{\parallel},i\omega+i\Omega)
	\chi^{-1}(q_{\perp}, 0, i\omega)=0$.
   Clearly, the middle and right panels are very similar, justifying
   the use of the factorization approximation when $K_F$ is
   the largest scale.
   }
   \label{fig:comparison}
\end{figure*}

The difference between these two cases is depicted in Fig.~\ref{fig:comparison},
where contours of constant energy are plotted in the momentum plane
for $d=2$.

Obviously, when $\Lambda \ll K_F$, the exact and factorized
contours are almost indistinguishable.  Only when $\Lambda \sim K_F$
does the curvature of the Fermi surface become apparent and the
factorization approximation break down.

The figure also illustrates the fact that when $\Lambda \ll K_F$, the most highly
peaked portions of the integrand occupy significant phase space where
$q_{\perp} \gg q_{\parallel}$ for fixed energy (i.e. on each contour).
This is so despite the fact that $\Lambda_{\perp} \sim \Lambda_{\parallel}$,
and is the justification for the neglect of $q_{\parallel}$ terms in the 
bosonic propagators.  At the same time, we neglect $q_{\perp}$
pieces of the fermionic propagators because $K_F$ is large.

A less graphical way to see the above is as follows. 
Because $\Lambda \ll K_F$, the $q_{\parallel}$ integration
is dominated by the range ($\pm \Lambda^2/K_F$, $\pm \Lambda$).
Over this range, 
the fermionic propagator can be approximated
\begin{eqnarray}
	G^{-1}(q_{\perp}, q_{\parallel},i\omega) 
	&=& \omega^{2/3} - v_F(q_{\parallel} + q_{\perp}^2/2K_F) \nonumber \\
	&\approx& \omega^{2/3} - v_Fq_{\parallel} .
\label{G_inv}
\end{eqnarray}

The fermionic propagator tells us that
the most important regions of the integrand are for $q_{\parallel} \sim \omega^{2/3}$.  
At the same time, the bosonic propagator is most highly peaked around
$q \sim \omega^{1/3}$.  This means that $q^2 \sim \omega^{2/3}$.
Since the pole of the fermion propagator will force $q_{\parallel}^2 \sim \omega^{4/3}$,
this means that the boson propagator must have 
$q_{\perp}^2 \sim \omega^{2/3} \gg q_{\parallel}^2 \sim \omega^{4/3}$,
and thus
\begin{eqnarray}
	\chi^{-1}(q_{\perp}, q_{\parallel},i\omega) 
	&\approx& q_{\perp}^2 + \gamma|\omega|/q_{\perp}
\label{chi_inv}
\end{eqnarray}
All these approximations become exact in the $N_\Lambda \rightarrow 
\infty$ limit. Eqs.~(\ref{G_inv},\ref{chi_inv}) ensure the 
factorization of the $q_{\parallel}$ and $q_{\perp}$ integrations.

\section{Non-analytic corrections}
An intriguing question for future studies is the effect of non-analytic
Fermi-liquid corrections. Such non-analytic corrections to susceptibility
and other physical properties already exist in a standard Fermi liquid
theory~\cite{Belitz_RMP_SI,Efremov08_SI}. In generic cases, such
non-analytic corrections are relatively small. Empirically, it is 
in general hard to observe such non-analytic corrections. Even in the 
case of the {\it quantum critical point} of a weak ferromagnetic system, 
the existence of an extensive critical regime controlled by the 
fixed point without taking into account the 
non-analytic-Fermi-liquid corrections is supported by experimental
observations~\cite{Smith_Nature_SI}. For the ferromagnetic phase of 
the Kondo lattice system we have considered, this effect is expected 
to be even smaller: due to the breakdown of the Kondo effect,
the amplitudes of the non-analytic correction terms will be normalized 
in terms of the bare Fermi energy of the conduction electrons as opposed
to a renormalized Kondo energy. More broadly, just like it is important
to establish the Fermi liquid fixed point before such non-analytic 
corrections are analyzed in detail, we have focused on the existence
of a small-Fermi-surface ferromagnetic fixed point.

%% file: masterFile.bbl
\begin{thebibliography}{50}

\bibitem{Coleman05}
Coleman, P. \& Schofield, A.~J.
Quantum criticality.
{\it Nature} {\bf 433}, 226-229 (2005).

\bibitem{Gegenwart08}
Gegenwart, P., Si, Q. \& Steglich, F.
Quantum criticality in heavy-fermion metals.
{\it Nature Phys.} {\bf 4}, 186-197 (2008).

\bibitem{HvL07}
von L\"{o}hneysen, H., Rosch, A., Vojta, M. 
\& W\"{o}lfle, P.
Fermi-liquid instabilities at magnetic quantum phase transitions.
{\it Rev. Mod. Phys.} {\bf 79}, 1015-1075 (2007).

\bibitem{Sullow99}
Sullow, S., Aronson, M.~C., Rainford, B.~D. \& Haen, P.
Doniach phase diagram, revisited: From ferromagnet to Fermi liquid in 
pressurized CeRu$_2$Ge$_2$.
{\it Phys. Rev. Lett.} {\bf 82}, 2963-2966 (1999).

\bibitem{Larrea05}
Larrea, J. {\it et al.}
Quantum critical behavior in a CePt ferromagnetic Kondo lattice.
{\it Phys. Rev. B} {\bf 72}, 035129 (2005).

\bibitem{Drotziger06}
Drotziger, S. {\it et al.}
Suppression of ferromagnetism in CeSi$_{1.81}$ under temperature and pressure.
{\it Phys. Rev. B} {\bf 73}, 214413 (2006).

\bibitem{Sidorov03}
Sidorov, V.~A.
{\it et al.},
Magnetic phase diagram of the ferromagnetic Kondo-lattice compound
CeAgSb$_{\rm 2}$ up to 80 kbar.
{\it Phys. Rev. B} {\bf 67}, 224419 (2003).

\bibitem{Bauer05}
Bauer, E. D. {\it et al.}
Non-Fermi-Liquid behavior within the ferromagnetic phase in 
URu$_{2-x}$Re$_x$Si$_2$.
{\it Phys. Rev. Lett.} {\bf 94}, 046401 (2005).

\bibitem{Butch09}
Butch, N. P. \& Maple, M.~B.
Evolution of critical scaling behavior near a ferromagnetic 
quantum phase transition.
{\it Phys. Rev. Lett.} {\bf 103}, 076404 (2009).

\bibitem{Krellner07}
Krellner, C. {\it et al.}
CeRuPO: A rare example of a ferromagnetic Kondo lattice.
{\it Phys. Rev. B} {\bf 76}, 104418 (2007).

\bibitem{Bauer06}
Bauer, E. D. {\it et al.}
Physical properties of the ferromagnetic heavy-fermion compound UIr$_2$Zn$_{20}$.
{\it Phys. Rev. B} {\bf 74}, 155118 (2006)

\bibitem{Saxena00}
Saxena, S. S. {\it et al.}
Superconductivity on the border of itinerant-electron ferromagnetism in 
UGe$_2$.
{\it Nature} {\bf 406}, 587-592 (2000).

\bibitem{Levy07}
Levy, F., Sheikin, I. \& Huxley, A.
Acute enhancement of the upper critical field for superconductivity
approaching a quantum critical point in URhGe.
{\it Nature Physics} {\bf 3}, 460-463 (2007).

\bibitem{King91}
King, C. A. \& Lonzarich, G. G.
Quasiparticle properties in ferromagnetic CeRu$_2$Ge$_2$.
{\it Physica B} {\bf 171}, 161-165 (1991).

\bibitem{Yamagami94}
Yamagami, H. \& Hasegawa, A.
Fermi surface of LaRu$_2$Ge$_2$ and CeRu$_2$Ge$_2$ within local-density band 
theory.
{\it J. Phys. Soc. Jpn.} {\bf 63}, 2290-2302 (1994).

\bibitem{Ikezawa97}
Ikezawa, H. {\it et al.}
Fermi surface properties of ferromagnetic CeRu$_2$Ge$_2$.
{\it Physica B} {\bf 237-238}, 210-211 (1997).


\bibitem{Hewson97}
Hewson, A. C.
{\it The Kondo Problem to Heavy Fermions}
(Cambridge University Press, Cambridge, 1997).

\bibitem{Oshikawa00}
Oshikawa, M.
Topological approach to Luttinger's theorem and the Fermi surface of a Kondo 
lattice.
{\it Phys. Rev. Lett.} {\bf 84}, 3370-3373 (2000).

\bibitem{Settai02}
Settai, R. {\it et al.}
A change of the Fermi surface in UGe$_2$ across the critical pressure.
{\it J. Phys.: Condens. Matter} {\bf 14}, L29-L36 (2002).

\bibitem{Huxley}
Huxley, A. {\it et al.}
The co-existence of superconductivity and ferromagnetism
in actinide compounds.
{\it J. Phys.: Condens. Matter} {\bf 15}, S1945-S1955 (2003).

\bibitem{Schofield03}
Sandeman, K. G., Lonzarich, G. G. \& Schofield, A.~J.
Ferromagnetic superconductivity driven by changing Fermi 
surface topology.
{\it Phys. Rev. Lett.} {\bf 90}, 167005 (2003).

\bibitem{Stoner38}
Stoner, E. C.
Collective electron ferromagnetism.
{\it Proc. R. Soc. London, Ser. A} {\bf 165}, 372-414 (1938).

\bibitem{Perkins07}
Perkins, N.~B., Iglesias, J.~R., Nunez-Regueiro,
M.~D. \& Coqblin, B.
Coexistence of ferromagnetism and Kondo effect in the underscreened
Kondo lattice.
{\it Euro. Phys. Lett.} {\bf 79}, 57006 (2007).

\bibitem{Nagaoka66}
Nagaoka, Y.
Ferromagnetism in a narrow, almost half-filled s band.
{\it Phys. Rev.} {\bf 147}, 392-405 (1966).

\bibitem{Sigrist92}
Sigrist, M., Tsunetsugu, H., Ueda, K.,
\& Rice, T.~M.
Ferromagnetism in the strong-coupling regime of the one-dimensional
Kondo-lattice model.
{\it Phys. Rev. B} {\bf 46}, 13838-13846 (1992).

\bibitem{Batista03}
Batista, C.~D., Bonca, J., \& Gubernatis, J.~E.
Itinerant ferromagnetism in the periodic Anderson model.
{\it Phys. Rev. B} {\bf 68}, 214430 (2003).

\bibitem{Yamamoto07}
Yamamoto, S. J. \& Si, Q.
Fermi surface and antiferromagnetism in the
Kondo lattice: An asymptotically exact solution in $d > 1$ dimensions.
{\it Phys. Rev. Lett.} {\bf 99}, 016401 (2007).

\bibitem{Altshuler94}
Altshuler, B.~L., Ioffe, L.~B. \& Millis, A.~J.
Low energy properties of fermions with singular interactions.
{\it Phys. Rev. B} {\bf 50}, 14048-14065 (1994)

\bibitem{Shankar90}
Shankar, R.
Renormalization-group approach to interacting fermions.
{\it Rev. Mod. Phys.} {\bf 66}, 129-192 (1994).

\bibitem{Read95}
Read, N. \& Sachdev, S.
Continuum quantum ferromagnets at finite temperature and
the quantum hall effect.
{\it Phys. Rev. Lett.} {\bf 75}, 3509-3512 (1995).

\bibitem{Polchinski93}
Polchinski, J.
Low energy dynamics of the spinon-gauge system.
{\it Nucl. Phys. B} {\bf 422}, 617-633 (1994).

\bibitem{Smith_Nature08}
Smith, R.~P. {\it et al.}
Marginal breakdown of the Fermi-liquid state on the border of 
metallic ferromagnetism.
{\it Nature} {\bf 455}, 1220-1223 (2008).

\bibitem{Belitz_RMP05}
Belitz, D., Kirkpatrick, T.~R., \& Vojta, T.
How generic scale invariance influences quantum and classical phase
transitions.
{\it Rev. Mod. Phys.} {\bf 77}, 579-632 (2005).

\bibitem{Ingersent98}
Gonzalez-Buxton, C. \& Ingersent, K.
Renormalization-group study of Anderson and Kondo impurities in gapless
Fermi systems.
{\it Phys. Rev. B} {\bf 57}, 14254-14293 (1998).

\bibitem{Withoff90}
Withoff, D. \& Fradkin, E.
Phase transitions in gapless Fermi systems with magnetic impurities.
{\it Phys. Rev. Lett.} {\bf 64}, 1835-1838 (1990).

\bibitem{Rainford96}
Rainford, B.~D., Neville, A.~J., Adroja, D.~T., Dakin, S.~J.
\& Murani, A.~P.
Low temperature excitations in CeRu$_{\rm 2}$Si$_{\rm 2-x}$Ge$_{\rm x}$.
{\it Physica~B} {\bf 223-224}, 163-165 (1996).

\bibitem{Paschen04}
Paschen, S. {\it et al.}
Hall-effect evolution across a heavy-fermion quantum critical point.
{\it Nature} {\bf 432}, 881-885 (2004).

\bibitem{Park08}
Park, T. {\it et al.}
Isotropic quantum scattering and unconventional superconductivity.
{\it Nature} {\bf 456}, 366-368 (2008).

\bibitem{Ueda75}
Ueda, K. \& Moriya, T.
Contribution of spin fluctuations to the electrical and thermal resistivities of
weakly and nearly ferromagnetic metals.
{\it J. Phys. Soc. Jpn.} {\bf 39}, 605 (1975).

\bibitem{Vollhardt98}
Vollhardt, D. {\it et al.}
Metallic ferromagnetism: Progress in our understanding
of an old strong-coupling problem.
{\it Adv. in Solid State Phys.} {\bf 38}, 383-396 (1999);
arXiv:cond-mat/9804112.

\bibitem{Cruz08}
de la Cruz, C. {\it et al.}
Magnetic order close to superconductivity in the iron-based layered
LaO$_{\rm 1-x}$F$_{\rm x}$FeAs systems.
{\it Nature} {\bf 453}, 899-902 (2008).


\end{thebibliography}

\begin{thebibliography}{99}

\bibitem{Pandey08_SI}
Pandey, S. {\it et al.}
Fermionic representation for the ferromagnetic Kondo lattice model:
Diagrammatic study of spin-charge coupling effects on magnon excitations.
{\it Phys. Rev. B} {\bf 77}, 134447 (2008).

\bibitem{Kapetanakis08_SI}
Kapetanakis, M.~D. \& Perakis, I.~E.
Magnetization relaxation and collective spin excitations in correlated
double-exchange ferromagnets.
{\it Phys. Rev. B} {\bf 78}, 155110 (2008).

\bibitem{Read95_SI} 
Read, N. \& Sachdev, S.
Continuum quantum ferromagnets at finite temperature and 
the quantum hall effect.
{\it Phys. Rev. Lett.} {\bf 75}, 3509-3512 (1995).

\bibitem{Wen88_SI} Wen. X.~G. \& Zee, A.
Spin waves and topological terms in the mean-field
theory of two-dimensional ferromagnets and antiferromagnets.
{\it Phys. Rev. Lett.} {\bf 61}, 1025-1028 (1988).

\bibitem{Yamamoto07_SI} 
Yamamoto, S. J. \& Si, Q.
Fermi surface and antiferromagnetism in the
Kondo lattice: an asymptotically exact solution in $d > 1$ dimensions.
{\it Phys. Rev. Lett.} {\bf 99}, 016401 (2007).

\bibitem{Shankar94_SI} 
Shankar, R.
Renormalization-group approach to interacting fermions.
{\it Rev. Mod. Phys.} {\bf 66}, 129-192 (1994).

\bibitem{Altshuler94_SI} 
Altshuler, B.~L., Ioffe, L.~B. \& Millis, A.~J.
Low energy properties of fermions with singular interactions.
{\it Phys. Rev. B} {\bf 50}, 14048-14065 (1994)

\bibitem{Yamamoto10_SI}
Yamamoto, S.~J., \& Si, Q.
Renormalization group for mixed fermion-boson systems.
{\it Phys. Rev. B} {\bf 81}, 205106 (2010).

\bibitem{Sachdev_book_SI} Sachdev, S.
{\it Quantum Phase Transitions}
(Cambridge University Press, Cambridge, 1999).

\bibitem{Tsvelik_book_SI} Tsvelik, A.
{\it Quantum Field Theory in Condensed Matter Physics}
(Cambridge University Press, Cambridge, 2003).

\bibitem{Holstein73_SI} 
Holstein, T., Norton, R. E. \& Pincus, P.
de Haas-van Alphen effect and the specific heat of an electron gas.
{\it Phys. Rev. B} {\bf 8}, 2649-2656 (1973).

\bibitem{Lee89_SI} 
Lee, P. A.
Gauge field, Aharonov-Bohm flux, and high-T$_c$ superconductivity.
{\it Phys. Rev. Lett.} {\bf 63}, 680-683 (1989).

\bibitem{Polchinski93_SI} 
Polchinski, J.
Low energy dynamics of the spinon-gauge system.
{\it Nucl. Phys. B} {\bf 422}, 617-633 (1994).

\bibitem{Rech06_SI} 
Rech, J., Pepin, C. \& Chubukov, A.~V.
Quantum critical behavior in itinerant electron systems: Eliashberg theory and instability 
of a ferromagnetic quantum critical point.
{\it Phys. Rev. B} {\bf 74}, 195126 (2006).

\bibitem{Efremov08_SI}
Efremov, D.~V., Betrouras, J.~J. \& Chubukov, A.~V.
Non-analytic behavior of 2D itinerant ferromagnet.
arXiv:0804.2736v1.

\bibitem{Oganesyan01_SI} 
Oganesyan, V., Fradkin, E. \& Kivelson, S. A.
Quantum theory of a nematic Fermi fluid.
{\it Phys. Rev. B} {\bf 74}, 195126 (2001).

\bibitem{Vekhter04_SI} Vekhter, I. \& Chubukov, A.~V.
Non-Fermi-Liquid behavior in itinerant antiferromagnets.
{\it Phys. Rev. Lett.} {\bf 93}, 016405 (2004).

\bibitem{Kim94_SI} 
Kim, Y.~B., Furusaki, A., Wen, X.~G., \& Lee, P.~A.
Gauge-invariant response functions of fermions coupled to a gauge field
{\it Phys. Rev. B} {\bf 50}, 17917 (1994).

\bibitem{AGD_SI} Abrikosov, A.~A., Gorkov, L.~P.  \& Dzyaloshinski, I.~E.
{\it Methods of Quantum Field Theory in Statistical Physics}
(Dover Publications, New York, 1975).

\bibitem{Abanov03_SI} 
Abanov, A., Chubukov, A.~V., \& Schmalian, J.
Quantum-critical theory of the spin-fermion model and its application 
to cuprates: normal state analysis.
{\it Adv. Phys.} {\bf 52}, 119Ð218 (2003).

\bibitem{Chubukov05_SI} 
Chubukov, A.~V., \& Schmalian, J.
Superconductivity due to massless boson exchange in the strong-coupling limit.
{\it Phys. Rev. B} {\bf 72}, 174520 (2005).

\bibitem{Chubukov04_SI} 
Chubukov, A.~V., Pepin, C., \& Rech, J.
Instability of the quantum critical point of itinerant ferromagnets
{\it Phys. Rev. Lett.} {\bf 92}, 147003 (2004).

\bibitem{Sedrakyan09_SI} 
Sedrakyan, T.~A., \& Chubukov, A.~V.
Fermionic propagators for 2D systems with singular interactions.
{\it Phys. Rev. B} {\bf 79}, 115129 (2009).

\bibitem{Belitz_RMP_SI}
Belitz, D., Kirkpatrick, T.~R., \& Vojta, T.
How generic scale invariance influences quantum and classical phase 
transitions.
{\it Rev. Mod. Phys.} {\bf 77}, 579-632 (2005).

\bibitem{Smith_Nature_SI}
Smith, R.~P. {\it et al.}
Marginal breakdown of the Fermi-liquid state on the border of 
metallic ferromagnetism.
{\it Nature} {\bf 455}, 1220-1223 (2008).

\end{thebibliography}
